\documentclass[entropy,article,accept,moreauthors,pdftex,10pt,a4paper]{mdpi} 

\firstpage{1} 
\pubvolume{xx}
\issuenum{1}
\articlenumber{5}
\pubyear{2018}
\copyrightyear{2018}

\history{Received: date; Accepted: date; Published: date}

\usepackage{amssymb}
\DeclareMathOperator{\Tr}{Tr}

\Title{Quantum Chaos and Quantum Randomness--- Paradigms of Entropy Production on the Smallest Scales}

\Author{Thomas Dittrich $^{1}$\orcidA{0000-0001-7416-5033}}
\AuthorNames{Thomas Dittrich}
\address{%
$^{1}$ \quad Departamento de F\'\i sica, Universidad Nacional de Colombia, Bogot\'a, Colombia; tdittrich@unal.edu.co}
\corres{tdittrich@unal.edu.co; Tel.: +57-1-3165000 ext.\ 10276}

\abstract{Quantum chaos is presented as a paradigm of information processing by dynamical
systems at the bottom of the range of phase-space scales. Starting with a brief review of classical
chaos as entropy flow from micro- to macro-scales, I argue that quantum chaos came as an
indispensable rectification, removing inconsistencies related to entropy in classical chaos:
Bottom-up information currents require an inexhaustible entropy production and a diverging
information density in phase space, reminiscent of Gibbs' paradox in Statistical Mechanics.
It is shown how a mere discretization of the state space of classical models already entails
phenomena similar to hallmarks of quantum chaos, and how the unitary time evolution
in a closed system directly implies the ``quantum death?? of classical chaos. 
As complementary evidence, I discuss quantum chaos under continuous measurement.
Here, the two-way exchange of information with a macroscopic apparatus opens
an inexhaustible source of entropy and lifts the limitations implied by unitary quantum
dynamics in closed systems. The infiltration of fresh entropy restores permanent chaotic
dynamics  in observed quantum systems. Could other instances of stochasticity
in quantum mechanics be interpreted in a similar guise? Where observed
quantum systems generate randomness, that is, produce entropy without discernible
source, could it have infiltrated from the macroscopic meter? This speculation is worked out
for the case of spin measurement.}


\begin{document}

\section{Introduction}\label{sec1}
With the advent of the first publications proposing the concept of deterministic chaos and
substantiating it with a novel tool, computer simulations, more was achieved than just a major
progress in fields such as weather and turbulence \cite{Lor63}. They suggested a radically new
view of stochastic phenomena in physics. Instead of subsuming them under a gross global
category such as``chance??or ``randomness'', the concept of chaos offered a detailed analysis
on basis of deterministic evolution equations, thus indicating an identifiable source of stochasticity
in macroscopic phenomena. A seminal insight, to be expounded in Sect.\ \ref{sec2}, that arose
as a spin-off of the study of deterministic chaos, was that the entropy produced by chaotic
systems emerges by amplifying structures, initially contained in the smallest scales,
to macroscopic visibility \cite{Sha81}.

Inspired and intrigued by this idea, researchers such as Giulio Casati and Boris Chirikov saw its
potential as a promising approach also towards the microscopic foundations of statistical
mechanics, thus accepting the challenge to extend chaos to quantum mechanics. In the same
spirit as those pioneering works on deterministic chaos, they applied standard quantization to
Hamiltonian models of classical chaos and solved the corresponding Schr\"odinger equation
numerically \cite{CCIF83}, again utilizing the powerful computing equipment available at that time.
What they obtained was a complete failure on first sight, but paved the way towards
a deeper understanding not only of classical chaos, but also of the principles
of quantum mechanics, concerning in particular the way information is processed
on atomic scales: In closed quantum systems, the entropy production characteristic of classical
chaos ceases after a finite time and gives way to a behaviour that is not only deterministic
but even repetitive, at least in a statistical sense, hence does not generate novelty any longer.
The ``quantum death of classical chaos'' will be illustrated in Sect.\ \ref{sec31}.

The present article recalls this development, drawing attention to a third decisive aspect that is able
to reconcile that striking discrepancy found between quantum and classical dynamics in closed
chaotic systems. To be sure, the gap separating quantum from classical physics can be bridged to
a certain extent by semiclassical approximations, which interpolate between the two descriptions,
albeit at the expense of conceptual consistency and transparency \cite{Ozo88,BB97}. Also in
the case of quantum chaos they provide valuable insight into the fingerprints classical chaos
leaves in quantum systems. A more fundamental cause contributing to that discrepancy,
however, lies in the closure of the models employed to study quantum chaos.
It excludes an aspect of classicality that is essential for the phenomena we observe on
the macroscopic level: No quantum system is perfectly isolated, or else we could
not even know of its existence.

The r\^ole of being coupled to a macroscopic environment first came into sight in other areas
where quantum mechanics appears incompatible with basic classical phenomena, such as in
particular dissipation \cite{FV63,CL81,LCD&87}. Here, even classically, irreversible behaviour
can only be reconciled with time-reversal invariant microscopic equations of motion if
a coupling to a reservoir with a macroscopic number of degrees of freedom
(or a quasi-continuous spectrum) is assumed. Quantum mechanically,
this coupling not only explains an irreversible loss of energy,
it leads to a second consequence, at least as fundamental as dissipative energy loss:
a loss of information, which becomes manifest as decoherence \cite{JZ85,JZ&03}.

In the context of quantum dissipation, decoherence could appear as secondary to the energy
loss, yet it is the central issue in another context where quantum behaviour resisted
a satisfactory interpretation for a long time: quantum measurement. The ``collapse of the
wavepacket'' remained an open problem even within the framework of unitary quantum mechanics,
till it could be traced back as well to the presence of a macroscopic environment, incorporated in
the measurement apparatus \cite{Zur81,Zur82,Zur83,Zur84,Zur91,Zur03}. As such, the collapse is
not an annoying side effect but plainly indispensable, to make sure that the measurement leaves 
a lasting record in the apparatus, thus becoming a fact in the sense of classical physics.
Since there is no dissipation involved in this case, quantum measurement became
a paradigm of decoherence induced by interaction and entanglement with an environment.

The same idea, that decoherence is a constituent aspect of classicality, proves fruitful in
the context of quantum chaos as well \cite{ZP94}. It forms an essential complement to
semiclassical approximations, in that it lifts the ``splendid isolation'', which inhibits a sustained
increase of entropy in closed quantum systems. Section \ref{sec32} elucidates how the coupling
to an environment restores the entropy production, constituent for deterministic chaos,
at least partially in classically chaotic quantum systems. Combining decoherence
with dissipation, other important facets of quantum chaos come into focus: It opens
the possibility to study quantum effects also in phenomena related to dissipative chaotic
dynamics, notably strange attractors, which, as fractals, are incompatible with uncertainty.

The insight guiding this article is that in the context of quantum chaos, the interaction with
an environment has a double-sided effect: It induces decoherence, as a loss of information, e.g., on
phases of the central quantum system, but also returns entropy from the environment to the
chaotic system \cite{UZ89,ZP94}, which then fuels its macroscopic entropy production. If
indeed there is a two-way traffic, an interchange of entropy between system and environment, this
principle, applied in turn to quantum measurement, has a tantalizing consequence: It suggests
that besides decoherence, besides the collapse of the wavepacket, also the randomness apparent
in the outcomes of quantum measurements could be traced back to the environment, could be
interpreted as a manifestation of entropy that infiltrates from the macroscopic apparatus.
This speculation is illustrated in Sect.\ \ref{sec4} for the emblematic case of spin measurement.
While Sections \ref{sec2} to \ref{sec3} largely have the character of reviews, complementing
the work of various authors with some original material, Sect.\ \ref{sec4} is a perspective,
it presents a project in progress at the time of writing this report.

\section{Classical chaos and information flows between micro- and macroscales}\label{sec2}

\subsection{Overview}\label{sec21}

The relationship between dynamics and information flows has been pointed out by mathematical
physicists, such as notably Kolmogorov, much before deterministic chaos was (re)discovered
in applied science, as is evident for example in the notion of Kolmogorov-Sinai entropy
\cite{LL83}. It measures the information production by a system with at least one positive
Lyapunov exponent and represents a central result of research on dynamical disorder
in microscopic systems, relevant primarily for statistical mechanics. For models of macroscopic
chaos, typically including dissipation, an interpretation as a phenomenon that has to do
with a directed information flow between scales came only much later. A seminal work
in that direction is the 1980 article by Robert Shaw \cite{Sha81}, where, in a detailed
discussion in information theoretic terms, the bottom-up information flow
related to chaos is contrasted with the top-down flow underlying dissipation. 

Shaw argues that the contraction of phase-space area in a dissipative system results in an
increasing loss of information on its initial state, if the state of the system is observed with a
given constant resolution. Conversely, later states can be determined to higher and higher
accuracy from measurements of the initial state. Chaotic systems show the opposite tendency:
Phase-space expansion, as consequence of exponentially diverging trajectories, allows to retrodict
the initial from the present state with increasing precision, while forecasting the final state
requires more and more precise measurements of the initial state as their separation
in time increases.

Chaotic systems therefore produce entropy, at a rate given by their Lyapunov exponents, as is
also reflected in the spreading of any initial distribution with a finite extension. The divergence
of trajectories also indicates the origin of this information: The chaotic flow amplifies details
of the initial distribution with an exponentially increasing magnification factor. If the state of
the system is observed with constant resolution, so that the total information on the present state is
bounded, the gain of information on small details is accompanied by a loss of information on the
largest scale, which impedes inverting the dynamics: Chaotic systems are \emph{globally} irreversible,
while the irreversibility of dissipative systems is a consequence of their loosing \emph{local}
information into ever smaller scales.

We achieve a more complete picture already by going to Hamiltonian systems, systems with
a phase space of even dimension. Their phase-space flow is symplectic, it conserves
phase-space area or volume, so that every expansion in some direction of phase space must
be compensated by contraction in another direction. In terms of information flows, this means that a
bottom-up current from small to large scales, corresponding to chaotic dynamics, will be accompanied
by an opposite current of the same magnitude, returning information to small scales. In the
framework of Hamiltonian dynamics, however, the top-down current is not related to dissipation,
it is not irreversible but to the contrary, complements the picture in such a way that all in all,
the time evolution becomes reversible.

A direct consequence of volume conservation by Hamiltonian flows is that Hamiltonian dynamics
also \emph{conserves entropy}, see Appendix \ref{clentrocons}. As is true for the underlying
conservation of volume, this invariance proves to be even more general than energy conservation
and applies, e.g., also to systems with a time-dependent external force where the total energy
is \emph{not} conserved. It indicates how to integrate dissipative systems in this more
comprehensive framework: Dissipation and other irreversible macroscopic phenomena
can be described within a Hamiltonian setting by going to models that include microscopic
degrees of freedom, typically as heat baths comprising an infinite number of freedoms,
on an equal footing in the equations of motion. In this way, entropy conservation extends
to the entire system. 

The conservation of the total entropy in systems comprising two or more degrees
of freedom or subsystems cannot be broken down, however, to a global sum rule
that would imply a simple exchange of information through currents among subsystems.
The reason is that in the presence of correlations, there exists a positive amount of mutual
information which prevents subdividing the total information content uniquely into contributions
associated to subsystems or individual degrees of freedom. Notwithstanding, if the partition
is not too complex, as is the case for a central system coupled to a thermal reservoir or heat bath,
it is still possible to keep track of internal information flows between these two sectors. For the
particular instance of dissipative chaos, a gross picture emerges that comprises three components:

\begin{itemize}[leftmargin=*,labelsep=5.8mm]
\item	a ``vertical'' current from large to small scales in certain dimensions within the central system,
representing the entropy loss that accompanies the dissipative loss of energy,
\item	an opposite vertical current, from small to large scales, induced by the chaotic dynamics in
other dimensions of the central system,
\item	a ``horizontal'' exchange of information between the central system and the heat bath,
including a redistribution of entropy within the reservoir, induced by its internal dynamics.
\end{itemize}

On balance, more entropy must be dumped by dissipation into the heat bath than is lifted
by chaos into the central system, thus maintaining consistency with the Second Law.
In phenomenological terms, this tendency is reflected in the overall contraction of a dissipative
chaotic system onto a strange attractor. After transients have faded out, the chaotic dynamics
then develops on a sub-manifold of reduced dimension of the phase space of the central system,
given by the attractor. For the global information flow it is clear that in a macroscopic chaotic
system, the entropy that surfaces at large scales by chaotic phase-space expansion has partially
been injected into the small scales from microscopic degrees of freedom of the environment.

Processes converting macroscopic structures into microscopic entropy, such as dissipation,
are the generic case. This report, however, is dedicated to the exceptional cases,
notably chaotic systems, which turn microscopic noise into macroscopic randomness.
The final section is intended to demonstrate that processes even belong to this
category where this is far less evident, in particular quantum measurements.

\subsection{Example 1: Bernoulli map and baker map} \label{clbaker}\label{sec22}

\begin{figure}[H]
\centering
\includegraphics[width=8 cm]{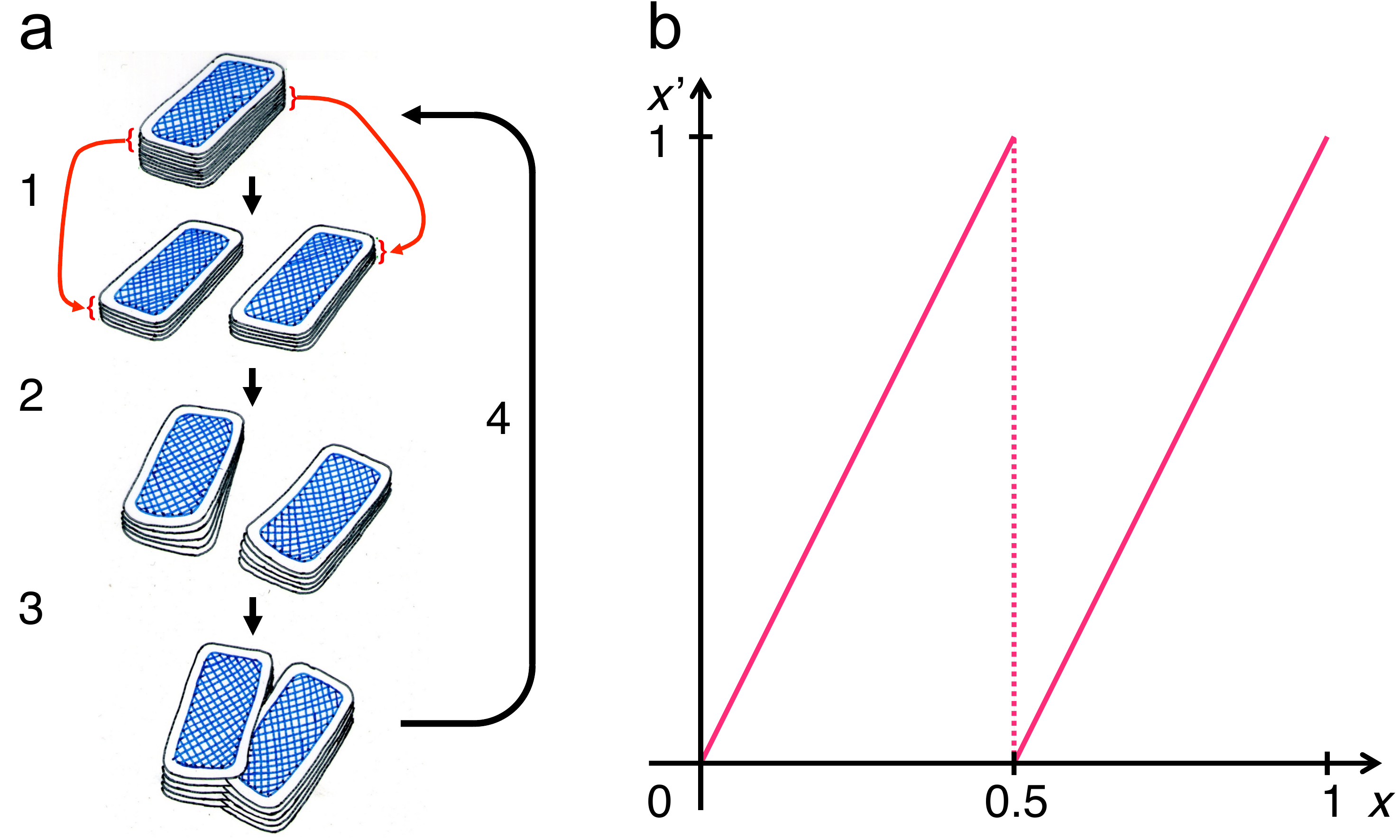}
\caption{The Bernoulli map can be understood as modelling a popular card shuffling
technique (\textbf{a}). It consists of three steps, (1) dividing the card deck into two halves
of equal size, (2) fanning the two half decks out to twice the original thickness, and (3)
intercalating one into the other as by the zipper method. (\textbf{b}) Replacing the
discrete card position in the deck by a continuous spatial coordinate, it reduces to
a map with a simple piecewise linear graph, cf.\ Eq.\ (\protect\ref{bernoullimap}).}
\label{figbernoulli}
\end{figure}

Arguably the simplest known model for classical deterministic chaos is the Bernoulli map
\cite{Sch84,Ott02}, a mapping of the unit interval onto itself that deviates from linearity only by
a single discontinuity. It is defined as 

\begin{equation} \label{bernoullimap}
x\mapsto x' = 2x \,({\rm mod}\, 1) = \begin{cases}
2x    & \text{$0 \leq x < 0.5$,} \\
2x-1 & \text{$0.5 \leq x < 1$,}
\end{cases}
\end{equation}

\noindent
and can be interpreted as a mathematical model of a popular card-shuffling technique (Fig.\
\protect\ref{figbernoulli}). The way it generates information by lifting it from scales
too small to be resolved to macroscopic visibility becomes immediately apparent
if the argument $x$ is represented as a binary sequence,
$x = \sum_{n=1}^{\infty} a_n 2^{-n}$, $a_n \in \{0,1\}$, so that map operates as

\begin{equation} \label{binbernoulli}
x' = 2\left(\sum_{n=1}^{\infty} a_n 2^{-n}\right) ({\rm mod}\, 1) = 
\sum_{n=1}^{\infty} a_n 2^{-n+1} ({\rm mod}\, 1) =
\sum_{n=1}^{\infty} a_{n+1} 2^{-n},
\end{equation}

\noindent
that is, the image $x'$ has the binary expansion

\begin{equation} \label{binbernoullires}
x' = \sum_{n=1}^{\infty} a'_n 2^{-n}, \quad \text{with}\; a'_n = a_{n+1}.
\end{equation}

\begin{figure}[H]
\centering
\includegraphics[width=4.5 cm]{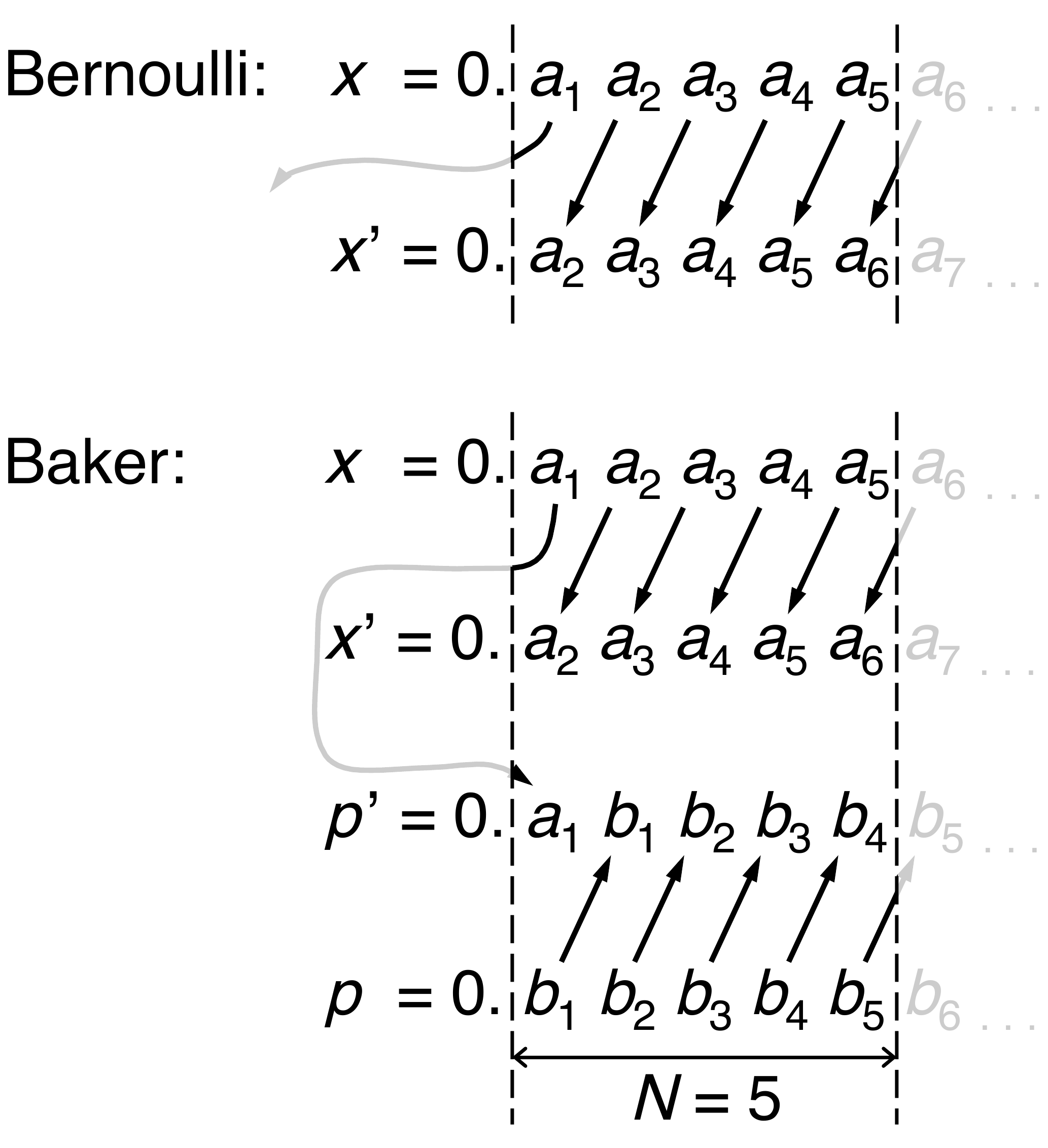}
\caption{Representing the Bernoulli map, Eq.\ (\protect\ref{bernoullimap}),
in terms of its action on a symbol string, the position encoded as a binary sequence,
see Eq.\ (\protect\ref{binbernoulli}), reveals  that it corresponds to a rigid shift by one symbol
of the string towards the most significant digit (upper panel). Encoding the baker map,
Eq.\ (\protect\ref{bakermap}), in the same way, Eq.\ (\protect\ref{binbaker}),
shows that the upward symbol shift in $x$ is complemented by a corresponding
downward shift in $p$ (lower panel). The loss of the most significant digit in
the Bernoulli map or its transfer from position to momentum in the baker map
are compensated by an equivalent gain or loss at the least significant digits, if
a constant finite resolution is taken into account, here limiting
the binary code to $N = 5$ digits (dashed vertical lines).}
\label{figbakernoulli}
\end{figure}

The action of the map consists in shifting the sequence of binary coefficients rigidly by one position
to the left (the ``Bernoulli shift'') and discarding the most significant digit $a_1$. In terms of
information, this operation creates exactly one bit per time step, coming from the smallest
resolvable scales, and at the same time looses one bit at the largest scale
(Fig.\ \protect\ref{figbakernoulli}a), which renders the map non-invertible.

\begin{figure}[H]
\centering
\includegraphics[width=9 cm]{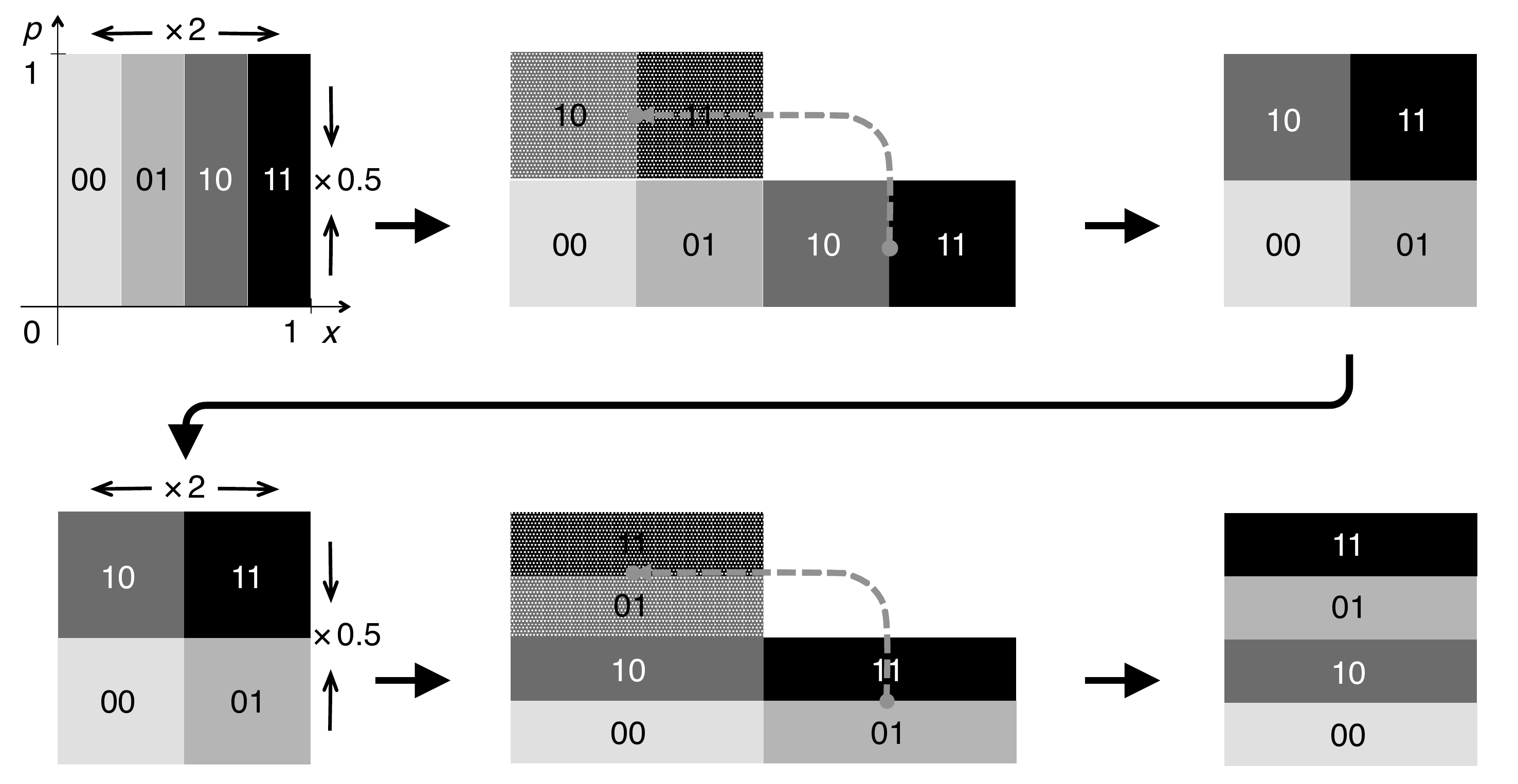}
\caption{
The baker map complements the Bernoulli map, Fig.\ \protect\ref{figbernoulli}, by a second
coordinate $p$, canonically conjugate to the position $x$, so as to become
consistent with the framework of symplectic phase-space geometry. Defining the map
for $p$ as the inverse of the Bernoulli map applied to $x$, a map of the unit square onto
itself results, see Eq.\ (\protect\ref{bakermap}), that is equivalent to a combination
of stretching and folding steps. The figure shows two subsequent applications
of the baker map and its effect on the binary code associated to a set of four
phase-space cells.}
\label{figbaker}
\end{figure}

By adding another dimension, the Bernoulli map is readily complemented so as to become
compatible with symplectic geometry. As the action of the map on the second coordinate,
say $p$, has to compensate for the expansion by a factor 2 in $x$, this suggests
modelling it as a map of the unit square onto itself, contracting $p$  by the same factor 2,

\begin{equation} \label{bakermap}
\begin{pmatrix}x\\p\end{pmatrix} \mapsto \begin{pmatrix}x'\\p'\end{pmatrix}, \quad
\begin{pmatrix}x'\\p'\end{pmatrix} = 
\begin{pmatrix}2x \,({\rm mod}\, 1) \\ \frac{1}{2} \bigl(p+{\rm int}(2x)\bigr)\end{pmatrix},
\end{equation}

\noindent
known as the baker map \cite{LL83,Ott02}. Geometrically, it can be interpreted as a combination
of stretching (by the expanding action of the Bernoulli map) and folding
(corresponding to the discontinuity of the Bernoulli map) (Fig.\ \protect\ref{figbaker}).
Being volume conserving, the baker map \emph{is} invertible. The inverse map reads

\begin{equation} \label{rekabmap}
\begin{pmatrix}x'\\p'\end{pmatrix} \mapsto \begin{pmatrix}x\\p\end{pmatrix}, \quad
\begin{pmatrix}x\\p\end{pmatrix} = 
\begin{pmatrix}\frac{1}{2} \bigl(x'+{\rm int}(2p')\bigr)\\2p \,({\rm mod}\, 1)\end{pmatrix}.
\end{equation}

\noindent
It interchanges the operations on $x$ and $p$ of the forward baker map.

The information flows underlying the baker map are revealed by encoding also $p$ as a
binary sequence, $p = \sum_{n=1}^{\infty} b_n 2^{-n}$. The action of the map again translates
to a rigid shift,

\begin{equation} \label{binbaker}
p' = \sum_{n=1}^{\infty} b'_n 2^{-n}, \quad \text{with}\; b'_n := \begin{cases}
a_1      & \text{$n = 1$,}\\
b_{n-1} & \text{$n \geq 2$.}
\end{cases}
\end{equation}

\noindent
However, it now moves the sequence by one step \emph{to the right}, that is, from
large to small scales. The most significant digit $b'_1$, which is not contained in the original
sequence for $p$, is transferred from the binary code for $x$, it recovers exactly
the coefficient $a_1$ that is discarded due to the expansion in $x$. This ``pasternoster
mechanism'' reflects the invertibility of the map. The upward information current
in $x$ is turned around to become a downward current in $p$
(Fig.\ \protect\ref{figbakernoulli}b). A full circle cannot be closed, however, as long as the
``depth'' from where and to which the information current reaches, remains unrestricted
by some finite resolution, indicated in Fig.\ \protect\ref{figbakernoulli}, as is manifest in
the infinite upper limit of the sums in Eqs.\ (\ref{binbernoulli},\ref{binbernoullires},\ref{binbaker}).

Generalizing the baker map so as to incorporate dissipation is straightforward
\cite{Sch84,Ott02}: Just insert a step that contracts phase space towards the origin
in the momentum direction, for example preceding the stretching and folding
operations of Eq.\ (\ref{bakermap}),

\begin{equation} \label{dissibakermap}
\begin{pmatrix}x\\p \end{pmatrix} \mapsto \begin{pmatrix}x'\\p' \end{pmatrix} = 
\begin{pmatrix}x \\ ap \end{pmatrix}, \quad
\begin{pmatrix}x'\\p'\end{pmatrix} \mapsto \begin{pmatrix}x''\\p'' \end{pmatrix}  = 
\begin{pmatrix}2x \,({\rm mod}\, 1) \\ \frac{1}{2} \bigl(p+{\rm int}(2x)\bigr)\end{pmatrix}.
\end{equation}

\begin{figure}[H]
\centering
\includegraphics[width=15.5 cm]{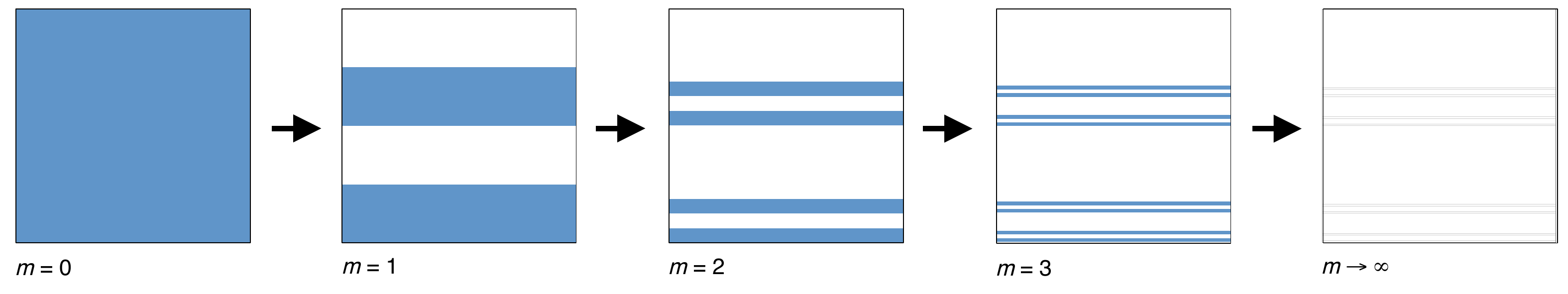}
\caption{A dissipative version of the baker map is created by preceding each iteration
of the map, as in Fig.\ \protect\ref{figbaker}, with a contraction by a factor $a$ in
$p$ (vertical axis), not compensated by a corresponding expansion in $x$ (horizontal axis),
see Eq.\ (\protect\ref{dissibakermap}). The figure illustrates this process for
a homogeneous initial density distribution ($m = 0$) and a contraction factor
$a = 0.5$ over the first three steps, $m = 1,\,2,\,3$. Asymptoticaly for $ m \to \infty$,
under the alternation of contraction and splitting, the distribution condenses
onto a strange attractor (rightmost panel) with a fractal dimension $ D = 1.5$.}
\label{figdissibaker}
\end{figure}

\noindent
A contraction by a factor $a$, $0 < a \leq 1$, models a dissipative reduction
of the momentum by the same factor. Figure \ref{figdissibaker} illustrates for the first
three steps how the generalized baker map operates, starting from a homogeneous
distribution over the unit square. For each step, the volume per strip reduces by
$a/2$ while the number of strips doubles, so that the overall volume reduction is
given by $a$. Asymptotically, a strange attractor emerges (rightmost panel in
Fig.\ \ref{figdissibaker}) with a fractal dimension, calculated as box-counting dimension
\cite{KS04},

\begin{equation} \label{dissibakerattrac}
D_0 = \frac{\text{log(volume contraction)}}{\text{log(scale factor)}} =
\frac{\ln(1/2)}{\ln(a/2)} = \frac{\ln(2)}{\ln(2) + \ln(1/a)}.
\end{equation}

\noindent
For example, for $a = 0.5$, as in Fig.\ \ref{figdissibaker}, a dimension $D_0 = 0.5$
results for the vertical cross section of the strange attractor, hence $D = 1.5$
for the entire manifold.

This model of dissipative chaos is simple enough to allow for a complete balance
of all information currents involved. Adopting the same binary coding
as in Eq.\ (\ref{binbaker}), a single dissipative step of the mapping, with $a = 0.5$,
(\ref{dissibakermap}) has the effect

\begin{equation} \label{bindissibaker}
p' = \frac{p}{2} = \frac{1}{2} \sum_{n=1}^{\infty} b'_n 2^{-n} =
\sum_{n=1}^{\infty} b'_n 2^{-n-1}.
\end{equation}

\noindent
That is, if $p$ is represented as $p = 0.\,b_1\,b_2\,b_3\,b_4\,\ldots$, $p'$ as
$p' = 0.\,b'_1\,b'_2\,b'_3\,b'_4\,\ldots$, the new binary coefficients are given by
a rigid shift by one unit to the right, but with the leftmost digit replaced by 0,

\begin{equation} \label{dissibakershift}
b'_n = \begin{cases}
0      & \text{$n = 1$,}\\
b_{n-1} & \text{$n \geq 2$.}
\end{cases}
\end{equation}

\begin{figure}[H]
\centering
\includegraphics[width=14 cm]{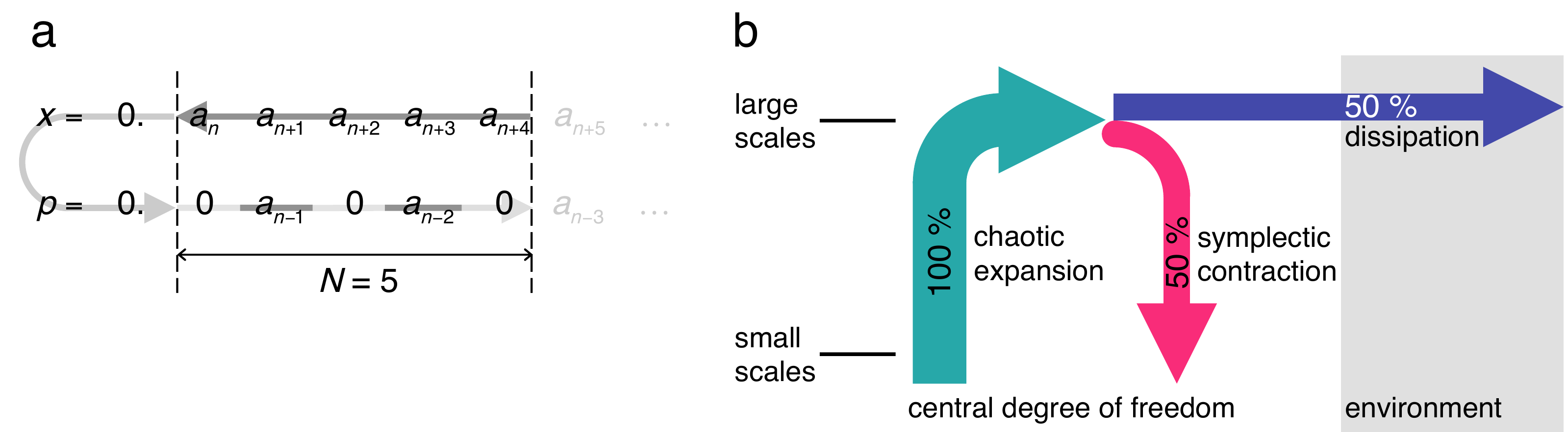}
\caption{(\textbf{a}) In terms of binary strings that encode position $x$ and
momentum $p$, resp., including dissipative contraction by a factor $a = 0.5$
in the baker map (see Fig.\ \protect\ref{figdissibaker}) results in an additional digit 0
fitted in between every two binary digits that are transferred from the upward Bernoulli shift in
$x$ to the downward shift in $p$. (\textbf{b}) Translated to bottom-up (green) and top-down
(pink) information currents, this means that half of the microscopic information
arriving at large scales by chaotic expansion is diverted by dissipation (blue)
to the environment, thus returning to small scales, but in adjacent degrees of freedom.}
\label{figdissibakerinfflow}
\end{figure}   

Combined with the original baker map (\ref{binbaker}), this additional step
fits in one digit 0 each between every two binary digits transferred from
position to momentum (Fig.\ \protect\ref{figdissibaker}). In terms of information currents,
this means that only half of the information lifted up by chaotic expansion in $x$ is returned
to small scales by the compensating contraction in $p$, the other half is diverted
by dissipation (Fig.\ \ref{figdissibakerinfflow}). This particularly simple picture
of course owes itself to the special choice of $a = 0.5$. Still, for other values of $a$,
different from $1/2$ or an integer power thereof, the situation will be qualitatively
the same. The fact that the dissipative information loss occurs here at the largest scales,
along with the volume conserving chaotic contraction in $p$, not at the smallest as would be 
expected on physical grounds, is an artefact of the utterly simplified model.

\subsection{Example 2: Kicked rotor and standard map}\label{sec23}

A model that comes much closer to an interpretation as a physical system than the Bernoulli
and baker maps is the kicked rotor \cite{Chi79,LL83,Ott02}. It can be motivated as an example,
reduced to a minimum of details, of a circle map, a discrete dynamical system conceived
to describe the phase-space flow in Hamiltonian systems close to integrability.
The kicked rotor, the version in continuous time of this model, can even be defined by a
Hamiltonian, but allowing for a time-dependent external force,

\begin{equation} \label{krhamil}
H(p,\theta,t) = \frac{p^2}{2} + V(\theta) \sum_{n=-\infty}^\infty \delta(t-n), \quad
V(\theta) = K \cos(\theta).
\end{equation}

\noindent
It can be interpreted as a plane rotor with angle $\theta$ and angular momentum $p$
and with unit inertia, driven by impulses that depend on the angle as a nonlinear function,
a pendulum potential, and on time as a periodic chain of delta kicks. Their strength
is controlled by the parameter $K$.

Reducing the continuous-time Hamiltonian (\ref{krhamil}) to a corresponding
discrete-time version in the form of a map is not a unique operation but depends,
for example, on the way stroboscopic time sections are inserted relative to the kicks.
For instance, if they follow immediately after each delta kick,
$t_n = \lim_{\epsilon\searrow 0^+} (n+\epsilon)$, $n\in \mathbb{Z}$,
the map from $t_n$ to $t_{n+1}$ reads

\begin{equation} \label{standardmap}
\begin{pmatrix}p \\ \theta\end{pmatrix} \mapsto \begin{pmatrix}p' \\ \theta'\end{pmatrix}, \quad
\begin{pmatrix}p' \\ \theta'\end{pmatrix} = 
\begin{pmatrix}p+ K \sin(\theta')\\ \theta + p\end{pmatrix}.
\end{equation}

\noindent
It is often referred to as the standard or Chirikov map \cite{Chi79,LL83,Ott02}.

The dynamical scenario of this model is by far richer than that of the Bernoulli and baker maps and
constitutes a prototypical example of the Kolmogorov-Arnol'd-Moser (KAM) theorem \cite{LL83}.
The parameter $K$ controls the deviation of the system from integrability. While for $K = 0$,
the kicked rotor is integrable, equivalent to an unperturbed circle map, increasing $K$ leads
through a complex sequence of mixed dynamics, with regular and chaotic phase-space regions
interweaving each other in an intricate fractal structure. For large values of $K$, roughly
given by $K \gtrsim 1$, almost all regular structures in phase space disappear and the dynamics
becomes purely chaotic. For the cylindrical phase space of the kicked rotor, $(p,\theta) \in
\mathbb{R} \otimes [0,2\pi[$, this means that the angle approaches a homogeneous distribution
over the circle, while the angular momentum spreads diffusively over the cylinder,
a case of deterministic diffusion, here induced by the randomizing action of the kicks.

For finite values of $K$, the spreading of the angular momentum does not yet follow
a simple diffusion law, owing to small non-chaotic islands in phase space \cite{Kar83}.
Asymptotically for $K \to \infty$, however, the angular momentum spreads diffusively,

\begin{equation} \label{krdifflaw}
\langle (p_n - \langle p\rangle )^2 \rangle = D(K)n
\end{equation}
with a diffusion constant
\begin{equation} \label{krdiffconst}
D(K) = \frac{K^2}{2}.
\end{equation}

\noindent
This regime is of particular interest in the present context, as it allows for a simple
estimate of the entropy production. In the kicked rotor, information currents cannot be
separated as neatly as in the baker map into a macro-micro flow in one coordinate and a
micro-macro flow in the other. The complex fractal phase-space structures imply that
these currents are organized differently in each point in phase space. Nevertheless,
some global features, relevant for the total entropy balance, can be extracted without going
to such detail.

Introduce a probability density in phase space that carries the full information available on the
state of the system,

\begin{equation} \label{rholthetadef}
\rho:\;  \mathbb{R} \otimes [0,2\pi[ \to \mathbb{R}^+,\;
\mathbb{R} \otimes [0,2\pi[\, \ni (p,\theta) \mapsto \rho(p,\theta) \in \mathbb{R}^+, \quad
\int_{-\infty}^\infty {\rm d} p \int_0^{2\pi} {\rm d} \theta\, \rho(p,\theta) = 1.
\end{equation}

\noindent
This density evolves deterministically according to Liouville's theorem \cite{Gol80,LL83}

\begin{equation} \label{rholiouvilleltheta}
\frac{\rm d}{{\rm d}t} \rho(p,\theta,t) =
\bigl\{ \rho(p,\theta,t),H(p,\theta,t) \bigr\} + \frac{\partial}{\partial t} \rho(p,\theta,t),
\end{equation}

\noindent
involving the Poisson bracket with the Hamiltonian (\ref{krhamil}).
In order to extract the overall entropy production from the detailed density $\rho(p,\theta,t)$,
some coarse graining is required. In the case of the kicked rotor, it offers itself to integrate
$\rho(p,\theta,t)$ over $\theta$, since the angular distribution rapidly approaches homogeneity, 
concealing microscopic information in fine details, while the diffusive spreading
in $p$ contains the most relevant large-scale structure. A time-dependent probability
density for the angular momentum alone is defined projecting by the full distribution along $\theta$,

\begin{equation} \label{rholtimedef}
\rho_p(p,t) := \int_0^{2\pi} {\rm d} \theta\, \rho(p,\theta,t), \quad
\int_{-\infty}^\infty {\rm d} p\, \rho_p(p,t) = 1.
\end{equation}

\noindent
Its time evolution is no longer given by Eq.\ (\ref{rholiouville}) but follows a
Fokker-Planck equation,

\begin{equation} \label{krrhodiffeq}
\frac{\partial}{\partial t} \rho_p(p,t) = D(K) \frac{\partial^2}{\partial p^2} \rho_p(p,t).
\end{equation}

\noindent
For a localized initial condition, $\rho(p,0) = \delta(p - p_0)$,
Eq.\ (\ref{krrhodiffeq}) it is solved for $t > 0$ by a Gaussian with a width that increases
linearly with time
\begin{equation} \label{krrhogauss}
\rho_p(p,t) = \frac{1}{\sqrt{2\pi}\sigma(t)}
\exp\left(- \frac{(p-p_0)^2}{2 \bigl(\sigma(t)\bigr)^2}\right), \quad \sigma(t) = D(K)t.
\end{equation}

\noindent
Define the total information content of the density $\rho_p(p,t)$ as

\begin{equation} \label{rhoinfo}
I(t) = -c \int_{-\infty}^\infty {\rm d} p\, \rho_p(p,t) \ln\bigl(d_p \rho_p(p,t)\bigr),
\end{equation}

\noindent
where $c$ is a constant fixing the units of information (e.g., $c = \log_2(e)$ for bits
and $c = k_{\rm B}$, the Boltzmann constant, for thermodynamic entropy) and $d_p$ denotes
the resolution of angular momentum measurements. The diffusive spreading given by
Eq.\ (\ref{krrhogauss}) corresponds to a total entropy growing as
\begin{equation} \label{krentprod}
I(t) = \frac{c}{2}\left[\ln\left(\frac{2\pi D(K)t}{d_p^2}\right) + 1\right],
\end{equation}

\noindent
hence to an entropy production rate of

\begin{equation} \label{krentprodrate}
\frac{{\rm d}}{{\rm d} t} I(t) = \frac{c}{2t}.
\end{equation}

\noindent
This positive rate decays with time, but only algebraically, that is, without a definite time scale.

Even if dissipation is not the central issue here, including it to illustrate a few relevant
aspects in the present context is in fact straightforward. On the level of the discrete-time
map, Eq.\ (\ref{standardmap}), a linear reduction of the angular momentum leads to
the dissipative standard map or Zaslavsky map \cite{Zas78,SW85},

\begin{equation} \label{disstandardmap}
\begin{pmatrix}p \\ \theta\end{pmatrix} \mapsto \begin{pmatrix}p' \\ \theta'\end{pmatrix}, \quad
\begin{pmatrix}p' \\ \theta'\end{pmatrix} = 
\begin{pmatrix}e^{-\lambda}p+ K \sin(\theta')\\ \theta + e^{-\lambda}p\end{pmatrix}.
\end{equation}

\noindent
The factor $\exp(-\lambda)$ results from integrating the equations of motion

\begin{equation} \label{disstandardnewton}
\begin{split}
\dot p &= - \lambda p + K \sin(\theta) \sum_{n=-\infty}^\infty \delta(t-n), \\
\dot \theta &= p.
\end{split}
\end{equation}

\noindent
The Fokker-Planck equation (\ref{krrhodiffeq}) has to be complemented accordingly
by a drift term $\sim \partial \rho_p(p,t) / \partial p$,

\begin{equation} \label{krrhofopl}
\frac{\partial}{\partial t} \rho_p(p,t) = (1-\lambda) \frac{\partial}{\partial p} \rho_p(p,t) +
\frac{\partial}{\partial p}\left(D(K) +
\bigl( (1-\lambda)p\bigr)^2 \right)  \frac{\partial}{\partial p} \rho_p(p,t).
\end{equation}

\noindent
In the chaotic regime $K \gtrsim 1$ of the conservative standard map, the dissipative
map (\ref{disstandardmap}) approaches a stationary state characterized by a strange
attractor, see, e.g., Refs. \cite{Zas78,SW85}.

\subsection{Anticipating quantum chaos: classical chaos on discrete spaces}\label{sec24}

Classical chaos can be understood as the manifestation of information currents that lift
microscopic details to macroscopic visibility \cite{Sha81}. Do they draw from an inexhaustible
information supply on ever smaller scales? The question bears on the existence
of an upper bound of the information density in phase space or other physically relevant
state spaces, or equivalently, on a fundamental limit of distinguishability, an issue
raised notably also by Gibbs' paradox \cite{Rei65}. Down to which minute difference
between their states will two physical systems remain distinct? The question
has already been answered implicitly above by keeping the number of binary digits in Eqs.\
(\ref{binbernoulli},\ref{binbernoullires},\ref{binbaker}) indefinite, in agreement with
the general attitude of classical mechanics not to introduce any absolute limit of
distinguishability.

A similar situation arises if chaotic maps are simulated on digital machines with finite
precision and/or finite memory capacity \cite{CP82,HW85,WH86,BR87}. 
In order to assess the consequences of discretizing the state space of a chaotic system, 
impose a finite resolution in Eqs.\ (\ref{binbernoulli},\ref{binbernoullires},\ref{binbaker}),
say $d_x = 1/J$, $J = 2^N$ with $N \in \mathbb{N}$, so that the sums over binary digits only
run up to $N$. This step is motivated, for example, by returning to the card-shuffling technique
quoted as inspiration for the Bernoulli map (Fig.\ \protect\ref{figbernoulli}). A finite number of
cards, say $J$, in the card deck, corresponding to a discretization of the coordinate $x$ into
steps of size $d_x > 0$,  will substantially alter the dynamics of the model.

More precisely, specify the discrete coordinate as

\begin{equation} \label{discbernoullix}
x_j = \frac{j-1}{J},\quad j = 1,\,2,\,3,\,\ldots,\,J,\quad J = 2^N,\quad N \in \mathbb{N},
\end{equation}

\noindent
with a binary code

\begin{equation} \label{discbernoullibinx}
x = \sum_{n=1}^N a_n 2^{-n}.
\end{equation}

\noindent
A density distribution over the discrete space $(x_1,x_2,\ldots,x_J)$ can now be written
as a $J$-dimensional vector

\begin{equation} \label{discbernoullivecx}
{\boldsymbol \rho} = (\rho_1,\rho_2,\rho_3,\ldots,\rho_J),\quad \rho_j \in \mathbb{R}^+,\quad
\sum_{j=1}^N \rho_j = 1,
\end{equation}

\noindent
so that the Bernoulli map takes the form of a $(J\times J)$-permutation matrix ${\bf B}_J$,

\begin{equation} \label{discbernoullirho}
{\boldsymbol \rho} \mapsto {\boldsymbol \rho}' = {\bf B} _J{\boldsymbol \rho}.
\end{equation}

\noindent
These matrices reproduce the graph of the Bernoulli map, Fig.\ \ref{figbernoulli}, but discretized
on a $(J\times J)$ square grid. Moreover, they incorporate a deterministic version of the
step of interlacing two partial card decks in the shuffling procedure, in an alternating sequence
resembling a zipper. For example, for $J = 8$, $N = 3$, the matrix reads

\begin{equation} \label{discbernoullimatrix}
{\bf B} _8 = \begin{pmatrix}
1 & 0 & 0 & 0 & 0 & 0 & 0 & 0 \\
0 & 0 & 0 & 0 & 1 & 0 & 0 & 0 \\
0 & 1 & 0 & 0 & 0 & 0 & 0 & 0 \\
0 & 0 & 0 & 0 & 0 & 1 & 0 & 0 \\
0 & 0 & 1 & 0 & 0 & 0 & 0 & 0 \\
0 & 0 & 0 & 0 & 0 & 0 & 1 & 0 \\
0 & 0 & 0 & 1 & 0 & 0 & 0 & 0 \\
0 & 0 & 0 & 0 & 0 & 0 & 0 & 1
\end{pmatrix}.
\end{equation}

\begin{figure}[H]
\centering
\includegraphics[width=12 cm]{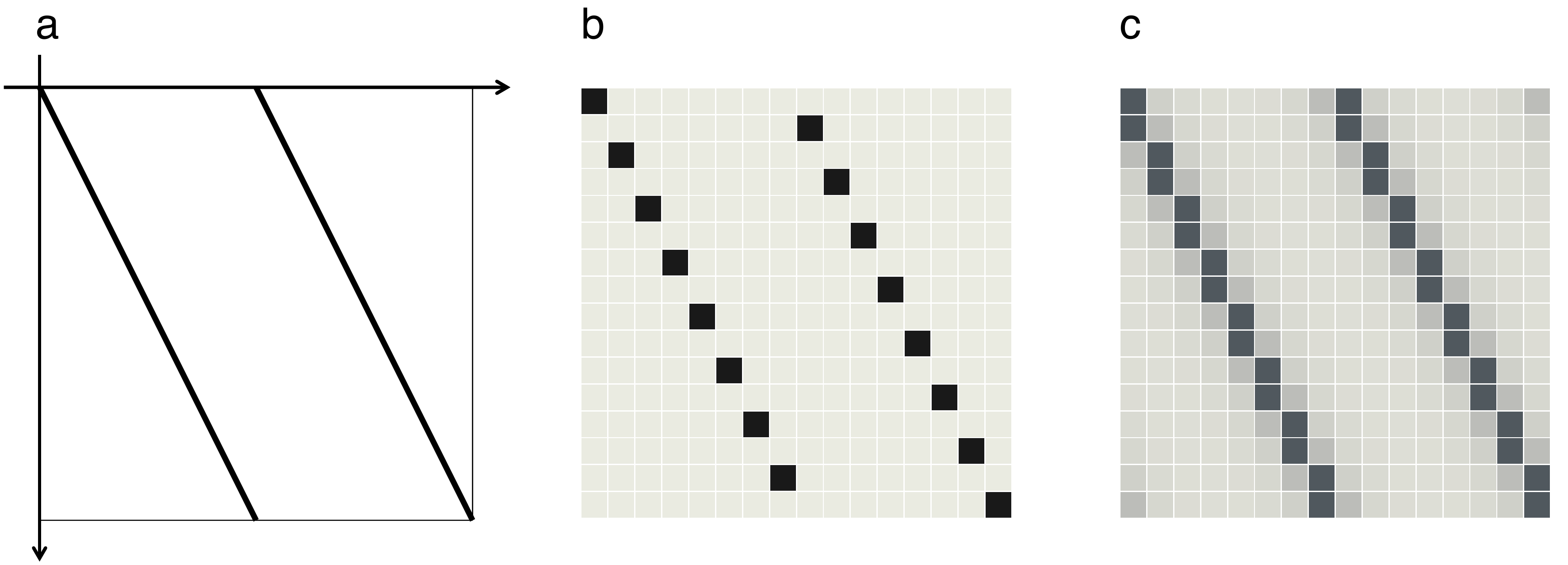}
\caption{Three versions of the Bernoulli map exhibit a common underlying structure.
The graph of the classical continuous map, Eq.\ (\protect\ref{bernoullimap}), panel
(\textbf{a}), recurs in the structure of the matrix generating the discretized
Bernoulli map (\textbf{b}), Eq.\ (\protect\ref{discbernoullimatrix}), here for cell number $J = 16$,
and becomes visible as well as marked ``ridges'' in the unitary transformation
generating (\textbf{c}) the quantum baker map, here depicted as
the absolute value of the transformation matrix in the position representation,
for a Hilbert space dimension $D_{\mathcal{H}} = J = 16$.
Grey-level code in (\textbf{b}) and (\textbf{c}) ranges from light grey (0) through black (1).}
\label{figpixelbaker}
\end{figure}

\noindent
The two sets of entries $= 1$ along slanted vertical lines represent the two branches
of the graph in Fig.\ \protect\ref{figbernoulli}, as shown in Fig.\ \protect\ref{figpixelbaker}b.

\begin{figure}[H]
\centering
\includegraphics[width=8 cm]{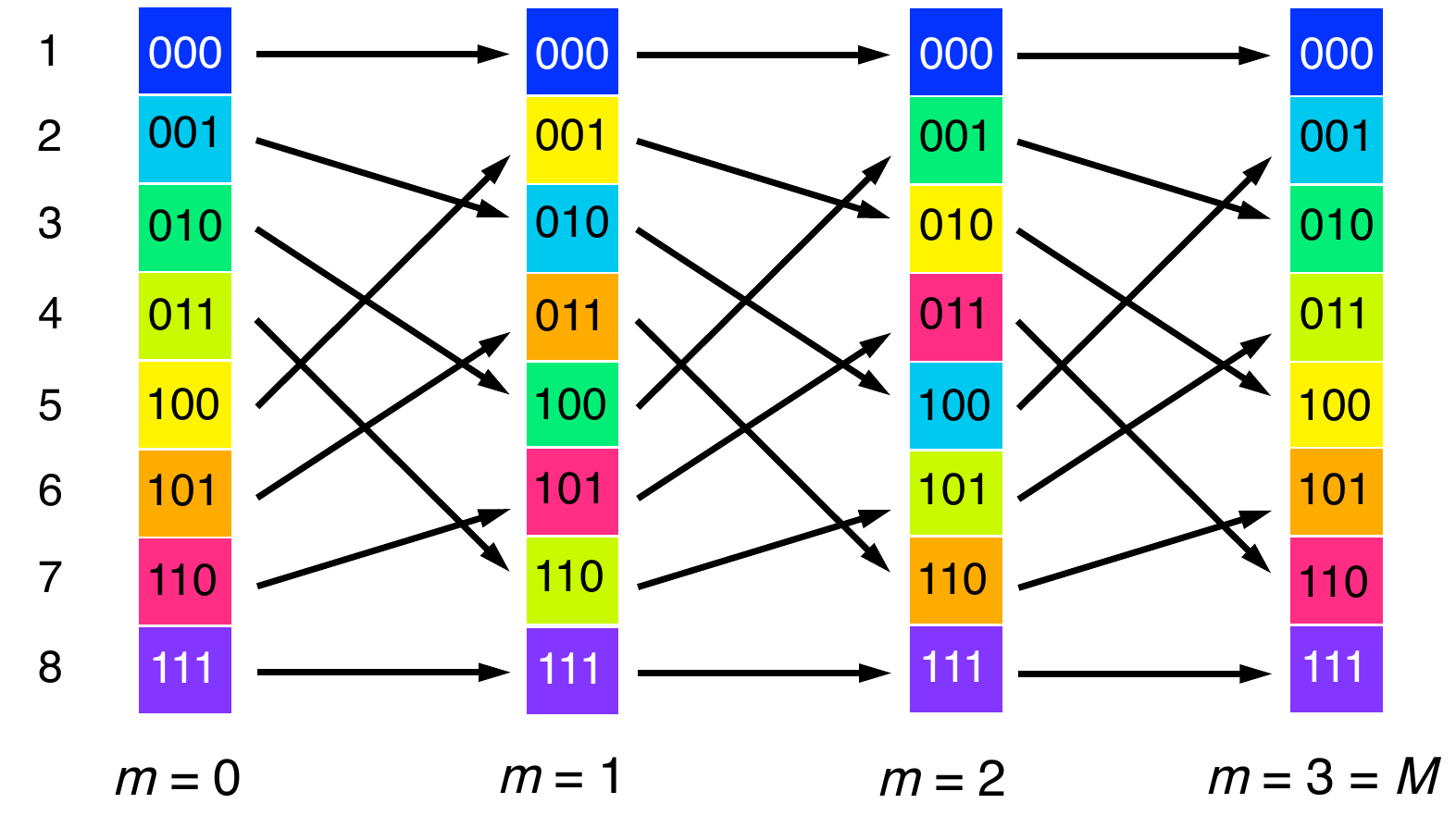}
\caption{If the discreteness of the cards in the card-shuffling model, see
Fig.\ \protect\ref{figbernoulli}a, is included in a corresponding discrete Bernoulli map,
it can be represented as a permutation matrix, Eq.\ (\protect\ref{discbernoullimatrix}).
The figure shows how it leads to a complete unshufling of the cards after a finite number
$M = {\rm lb}(J)$ of steps, here for $M = 3$. Moreover, a binary coding of the cell index
reveals that subsequent positions of a card are given by a permutation of the binary code.}
\label{figbernoulli8}
\end{figure}

A deterministic dynamics on a discrete state space comprising a finite number of states
must repeat after a finite number $M$ of steps, not larger than the total number of states.
In the case of the Bernoulli map, the recursion time is easy to calculate: In binary digits,
the position discretized to $2^N$ bins is specified by a sequence of $N$ binary coefficients
$a_n$. The Bernoulli shift moves this entire sequence in $M = N = {\rm lb}(J)$ steps, which is
the period of the map. Exactly how the reshuffling of the cards leads to the full recovery
of the initial state after $M$ steps is illustrated in Fig.\ \protect\ref{figbernoulli8}. That is, the
shuffling undoes itself after $M$ repetitions!

A similar, but even more striking situation occurs for the baker map, discretized in the same
fashion. While the $x$-component is identical to the discrete Bernoulli map, the $p$-component
is construed as inverse of the $x$-component, cf. Eq.\ (\ref{rekabmap}). Defining a
matrix of probabilities on the discrete $(J\times J)$ square grid that replaces
the continuous phase space of the baker map,

\begin{equation} \label{discbakerrho}
{\boldsymbol \rho}:\; \{1,\,\ldots,\,J\} \otimes \{1,\,\ldots,\,J\} \to \mathbb{R}^+,\;
(n,m) \mapsto \rho_{n,m},\quad \sum_{n,m=1}^J \rho_{n,m} = 1,
\end{equation}

\noindent
the discrete map takes the form of a similarity transformation,

\begin{equation} \label{discbakermap}
{\boldsymbol \rho} \mapsto {\boldsymbol \rho}' =
{\bf B}_J^{-1} {\boldsymbol \rho} {\bf B}_J^{\rm t} = {\bf B}_J^{\rm t} 
{\boldsymbol \rho} {\bf B}_J^{\rm t}.
\end{equation}

\noindent
The inverse matrix ${\bf B}_J^{-1}$ is readily obtained as the transpose of ${\bf B}_J$.
For example, for $N = 3$, it reads

\begin{equation} \label{discinvbernoullimatrix}
{\bf B} _8^{-1} = {\bf B} _8^{\rm t}  = \begin{pmatrix}
1 & 0 & 0 & 0 & 0 & 0 & 0 & 0 \\
0 & 0 & 1 & 0 & 0 & 0 & 0 & 0 \\
0 & 0 & 0 & 0 & 1 & 0 & 0 & 0 \\
0 & 0 & 0 & 0 & 0 & 0 & 1 & 0 \\
0 & 1 & 0 & 0 & 0 & 0 & 0 & 0 \\
0 & 0 & 0 & 1 & 0 & 0 & 0 & 0 \\
0 & 0 & 0 & 0 & 0 & 1 & 0 & 0 \\
0 & 0 & 0 & 0 & 0 & 0 & 0 & 1
\end{pmatrix}.
\end{equation}

\noindent
As for the forward discrete map, it resembles the corresponding continuous graph
(Fig.\ \protect\ref{figpixelbaker}a), with entries 1 now aligned along two slanted
horizontal lines (Fig.\ \protect\ref{figpixelbaker}b) .

\begin{figure}[H]
\centering
\includegraphics[width=15.9 cm]{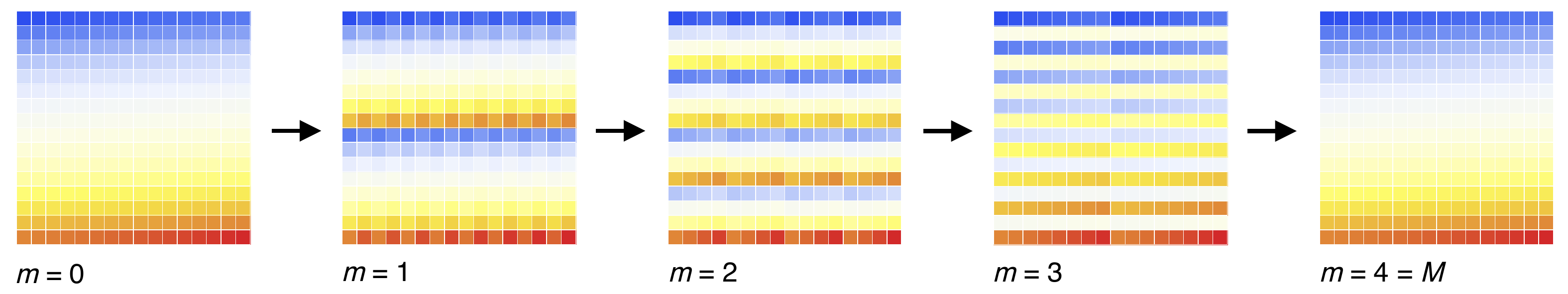}
\caption{The recurrence in the discrete Bernoulli map, see Fig.\ \protect\ref{figbernoulli8},
occurs likewise in the discrete baker map, Eq.\ (\protect\ref{discbakermap}). The figure
shows how the simultaneous expansion in $x$ (horizontal axis) and contraction in $p$
(vertical axis), in the pixelated two-dimensional state space entail an exact reconstruction
of the initial state, here after $M = {\rm lb}(16) = 4$ iterations of the map.}
\label{figbaker16x16}
\end{figure}   

Both the upward shift of binary digits of the $x$-component and the downward shift
of binary digits encoding $p$ now become periodic with period $M = N$, as for the discrete
baker map. The two opposing information currents thus close to a circle, resembling
a paternoster lift with a lower turning point at the least significant and an upper turning point
at the most significant digit (Fig.\ \protect\ref{figbaker16x16}).

The fate of deterministic classical chaos in systems comprising only a finite number
of discrete states (of a ``granular phase space'') has been studied in various systems
\cite{CP82,HW85,WH86,BR87}, with the same general conclusion that chaotic entropy production
gives way to periodic behaviour with a period determined by the size of the discrete
state space, that is, by the finite precision underlying its discretization.  To a certain extent,
this classical phenomenon anticipates the effects of quantization on chaotic dynamics,
but it provides at most a caricature of quantum chaos. It takes only a single, if crucial, tenet of
quantum mechanics into account, the fundamental bound uncertainty imposes
on the storage density of information in phase space, leaving all other
principles of quantum mechanics aside. Yet it anticipates a central feature of
quantum chaos, the repetitive character it attains in closed systems, and
it suggests how to interpret this phenomenon in terms of information flows.

\section{Quantum death and incoherent resurrection of classical chaos}\label{sec3}

While the ``poor man's quantization'' discussed in the previous section indicates
qualitatively what to expect if chaos is discretized, reconstructing classically chaotic
systems systematically in the framework of quantum mechanics also allows for a much more
profound analysis how these systems process information. It turns out that
quantum mechanics directs our view more specifically to the aspect of closure of
dynamical systems. Chaotic systems provide a particularly sensitive probe, more so
than systems with a regular classical mechanics, of the effects of a complete elimination of
external sources of entropy, since they react even to a weak infiltration of entropy
from the environment by a particularly drastic change of their dynamical behaviour.

\subsection{Quantum chaos in closed systems}\label{sec31}

A straightforward strategy to study the effect first principles of quantum mechanics
have on chaotic dynamics is quantizing models of classical chaos. This requires these
models, however, to be furnished with a minimum of mathematical structure, required to apply
basic elements of a quantum mechanical description. In essence, systems with a
volume conserving flow, generated by a Hamiltonian, on an even-dimensional state space
can be readily quantized. In the following, basic consequences of quantizing chaos
will be exemplified applying this strategy to the baker map and the kicked rotor.

\subsubsection{The quantized baker map} \label{sec311}

The baker map introduced in subsection \ref{clbaker} is an ideal model to consider
quantum chaos in a minimalist setting. It already comprises a coordinate together with its
canonically conjugate momentum and can be quantized in an elegant fashion
\cite{BV87,BV89,Sar90}. Starting from the operators $\hat x$ and $\hat p$,
$\hat p = -{\rm i}\hbar {\rm d}/{\rm d} x$ in the position representation,
with commutator $[\hat x,\hat p] = {\rm i}\hbar$, their eigenspaces are constructed as

\begin{equation} \label{xpeigensys}
\hat x \vert x\rangle = x \vert x\rangle,\; \hat p \vert p\rangle = p\vert p\rangle,\;
\langle x | p \rangle = \frac{{\rm e}^{{\rm i}px/\hbar}}{\sqrt{2\pi\hbar}}.
\end{equation}

\noindent
The finite phase space $[0,1[ \otimes [0,1[\, \subset\! \mathbb{R}^2$ can be imposed
on this pair of operators by assuming periodicity, say with period 1, both in $x$ and in $p$.
Periodicity in $x$ entails quantization of $p$ and vice versa, so that together, a
Hilbert space of finite dimension $J$ results, and the pair of eigenspaces (\ref{xpeigensys}) 
is replaced by

\begin{equation} \label{xpdisceigensys}
\hat x \vert j\rangle = \frac{j}{J} | j \rangle,\;
\hat p \vert l \rangle = \hbar l \vert l \rangle,\;
j,\,l = 0,\,\ldots,\, J-1,\;
\langle j \vert l \rangle = \frac{1}{\sqrt{J}} \, {\rm e}^{2\pi{\rm i}\,j l / J} =
(F_J)_{j,l},
\end{equation}

\noindent
that is, the transformation between the two spaces coincides with the discrete
Fourier transform, given by the $(J\times J)$-matrix $F_J$.

This construction facilitates the quantization of the baker map enormously. If we
phrase the classical map qualitatively as the sequence of operations
\begin{enumerate}
\item expand the unit square $[0,1[\, \otimes [0,1[\,$ by a factor 2 in $x$,
\item divide the expanded $x$-interval into two equal sections, $[0,1[$ and $[1,2[$,
\item shift the right one of the two rectangles (Fig.\ \protect\ref{figbaker}),
$(x,p) \in [1,2[\, \otimes\, [0,1[\,$, by 1 to the left in $x$ and by 1 up in $p$,
$[1,2[\, \otimes\, [0,1[\, \mapsto [0,1[\, \otimes\, [1,2[$,
\item contract by 2 in $p$,
\end{enumerate}
it translates to the following operations on the Hilbert space defined in
Eq.\ (\ref{xpdisceigensys}), assuming the Hilbert-space dimension $J$ to be even,
\begin{enumerate}
\item in the $x$-representation, divide the vector of coefficients
$(a_0,\,\ldots,\,a_{J-1})$, $\vert x\rangle = \sum_{j=0}^{J-1} a_j \vert j\rangle$,
into two halves, $(a_0,\,\ldots,\,a_{J/2-1})$ and $(a_{J/2},\,\ldots,\,a_{J-1})$
with indices running from 0 to $J/2-1$ and from $J/2$ to $J-1$, resp.,
\item transform both partial vectors separately to the $p$-representation, applying a
$(\frac{J}{2} \times \frac{J}{2})$-Fourier transform to each of them,
\item stack the Fourier transformed right half column vector on top of the Fourier transformed
left half, so as to represent the upper half of the spectrum of spatial frequencies,
\item transform the combined state vector in the $J$-dimensional momentum
Hilbert space back to the $x$ representation, applying an inverse $(J\times J)$-Fourier transform.
\end{enumerate}
All in all, this sequence of operations combines to a single unitary transformation matrix
in the position representation

\begin{equation} \label{qmbakerxtrans}
B_J^{(x)}= F_J^{-1} \begin{pmatrix} F_{J/2} & 0 \\ 0 & F_{J/2} \end{pmatrix}.
\end{equation}

Like this, it already represents a very compact quantum version of the Baker map
\cite{BV87,BV89}. It still bears one weakness, however: The origin $(j,l) = (0,0)$ of
the quantum position-momentum space, coinciding with the classical origin
$(x,p) = (0,0)$ of phase space, creates an asymmetry, as the diagonally opposite corner
$\frac{1}{J}(j,l) = \frac{1}{J}(J-1,J-1) = (1-\frac{1}{J},1-\frac{1}{J})$ does \emph{not} coincide with
$(x,p) = (1,1)$. In particular, it breaks the symmetry $x \to 1-x$, $p \to 1-p$ of the classical map.
This symmetry can be recovered on the quantum side by a slight modification \cite{Sar90} of
the discrete Fourier transform mediating between position and momentum representation,
a shift by $\frac{1}{2}$ of the two discrete grids. It replaces $F_J$ by

\begin{equation} \label{fouriersym}
\langle j \vert l \rangle = \frac{1}{\sqrt{J}}
\exp\left(2\pi{\rm i}\left(j+\frac{1}{2}\right)\left(l+\frac{1}{2}\right)\right) =:
(G_J)_{j,l},
\end{equation}

\noindent
and likewise for $F_{J/2}$. The quantum baker map in position representation
becomes accordingly

\begin{equation} \label{qmbakersymxtrans}
B_J^{(x)} = G_J^{-1} \begin{pmatrix} G_{J/2} & 0 \\ 0 & G_{J/2} \end{pmatrix}.
\end{equation}

\noindent
In momentum representation, it reads

\begin{equation} \label{qmbakersymptrans}
B_J^{(p)} = G_J B_J^{(x)} G_J^{-1} =
\begin{pmatrix} G_{J/2} & 0 \\ 0 & G_{J/2} \end{pmatrix} G_J^{-1} .
\end{equation}

\noindent
The matrix $B_J^{(x)}$ exhibits the same basic structure as its classical counterpart,
the $x$-component of the discrete baker map (\ref{discbernoullimatrix}),
but replaces the sharp ``crests'' along the graph of the original mapping by
smooth maxima (Fig.\ \protect\ref{figpixelbaker}c). Moreover, its entries are now complex.
In momentum representation, the matrix $B_J^{(p}$ correspondingly resembles
the $p$-component of the discrete baker map.

\begin{figure}[H]
\centering
\includegraphics[width=12 cm]{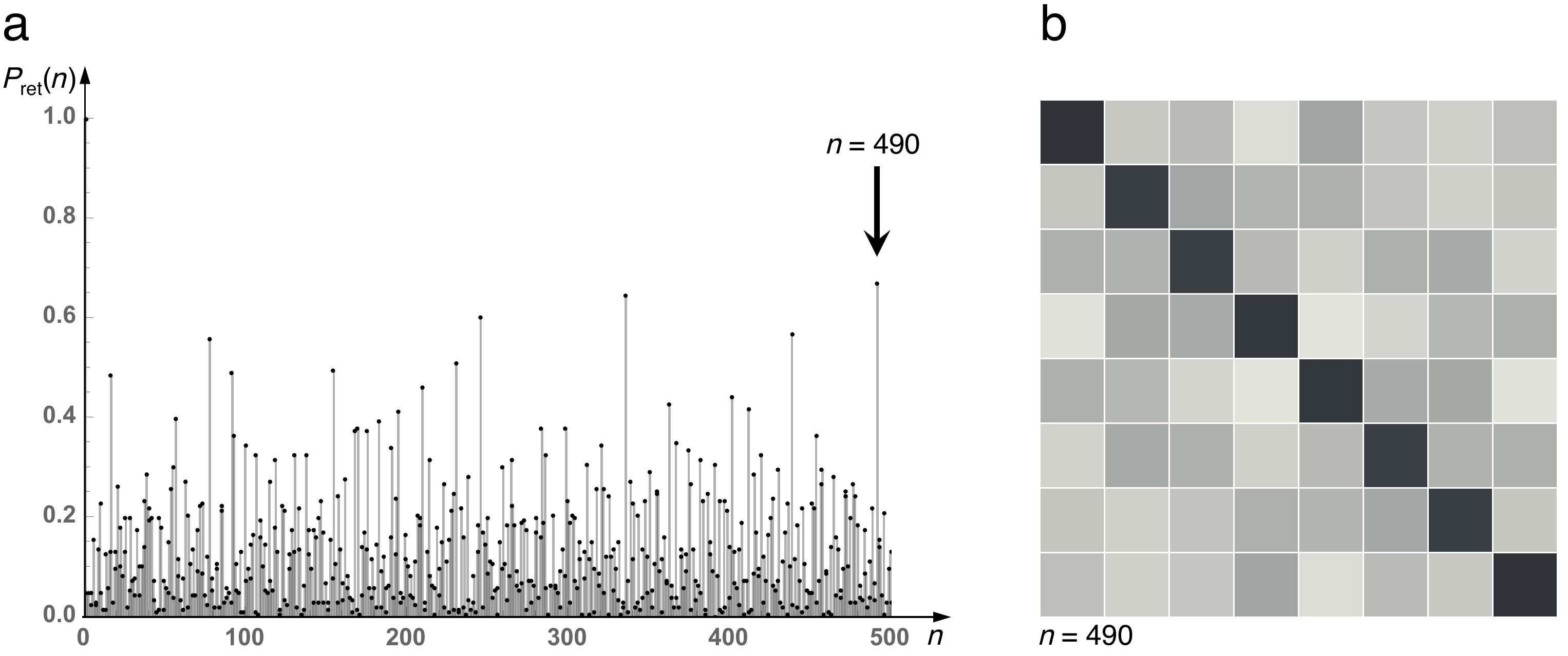}
\caption{Recurrences in the quantum baker map are neither periodic nor precise, as in
the classical version, see Fig.\ \protect\ref{figbaker16x16}, but occur as approximate revivals. They can be identified as marked peaks (\textbf{a}) of the return probability,
Eq.\ (\protect\ref{probret}). For the strong peak at time $n = 490$ (arrow in panel (\textbf{a})),
the transformation matrix in the position representation $B_{x,J}^n$ (\textbf{b}), cf.\
Eq.\ (\protect\ref{qmbakersymxtrans}), here with $J = 8$, indeed comes close to a unit matrix.
Grey-level code in (b) ranges from light grey (0) through black (1).}
\label{figqmbakerret}
\end{figure}

While the discretized classical baker map (\ref{discbakermap}) merely permutes the elements
of the classical phase-space distribution, the quantum baker map rotates complex
state vectors in a Hilbert space of finite dimension $J$. We cannot expect periodic exact
revivals as for the classical discretization. Instead, the quantum map is quasi-periodic, owing
to phases $\epsilon_n$ of its unimodular eigenvalues $e^{{\rm i}\epsilon_n}$, which in general
are not commensurate. With a spectrum comprising a finite number of discrete frequencies,
the quantum baker map therefore exhibits an irregular sequence of approximate revivals.
They can be visualized by recording the return probability,

\begin{equation} \label{probret}
P_{\rm ret}(n) = \left\vert \Tr [\hat U^n ] \right\vert^2
\end{equation}

\noindent
with the one-step unitary evolution operator $\langle j |\hat U | j'\rangle = (B_J^{(x)})_{j,j'}$.
Figure \ref{figqmbakerret}a shows the return probability of the $(8\times 8)$
quantum baker map for the first 500 time steps. Several near-revivals are visible;
the figure also shows the unitary transformation matrix $(B_J^{(x)})^n$
for $n = 490$ where it comes close to the $(8\times 8)$ unit matrix
(Fig.\ \protect\ref{figqmbakerret}b). Even with these caveats, it is evident that there
is no exponential decay of the return probability, as expected for classical chaos \cite{Sch84}.

\subsubsection{The quantum kicked rotor} \label{sec312}

By contrast to mathematical toy models such as the baker map, the kicked rotor
allows to include most of the features of a fully-fledged Hamiltonian dynamical
system, also in its quantization. With the Hamiltonian (\ref{krhamil}), a unitary
time-evolution operator over a single period of the driving is readily construed
\cite{CCIF83,She83}. Placing, as for the classical map, time sections immediately
after each kick, the time-evolution operator reads

\begin{equation} \label{qkrevop}
\hat U_{\rm QKR} = \hat U_{\rm kick} \hat U_{\rm rot},\quad
\hat U_{\rm kick} = \exp\left(-{\rm i}k\cos(\hat\theta)\right),\;
\hat U_{\rm rot} = \exp\left(-{\rm i}\hbar \hat l^2\right).
\end{equation}

\noindent
The parameter $k$ relates to the classical kick strength as $k = K/\hbar$.
Angular momentum $\hat p$ and angle $\hat\theta$ are now operators canonically
conjugate to one another, with commutator $[\hat p,\hat\theta] = -{\rm i}\hbar$.
The Hilbert space pertaining to this model is of infinite dimension, spanned for
example by the eigenstates of $\hat p$,

\begin{equation} \label{qkrhilbertspace}
\hat p \vert l \rangle = \hbar l | l \rangle,\; l \in \mathbb{Z},\;
\langle\theta | l \rangle = \frac{1}{\sqrt{2\pi\hbar}} \exp({\rm i}l\theta).
\end{equation}

\begin{figure}[H]
\centering
\includegraphics[width = 8 cm]{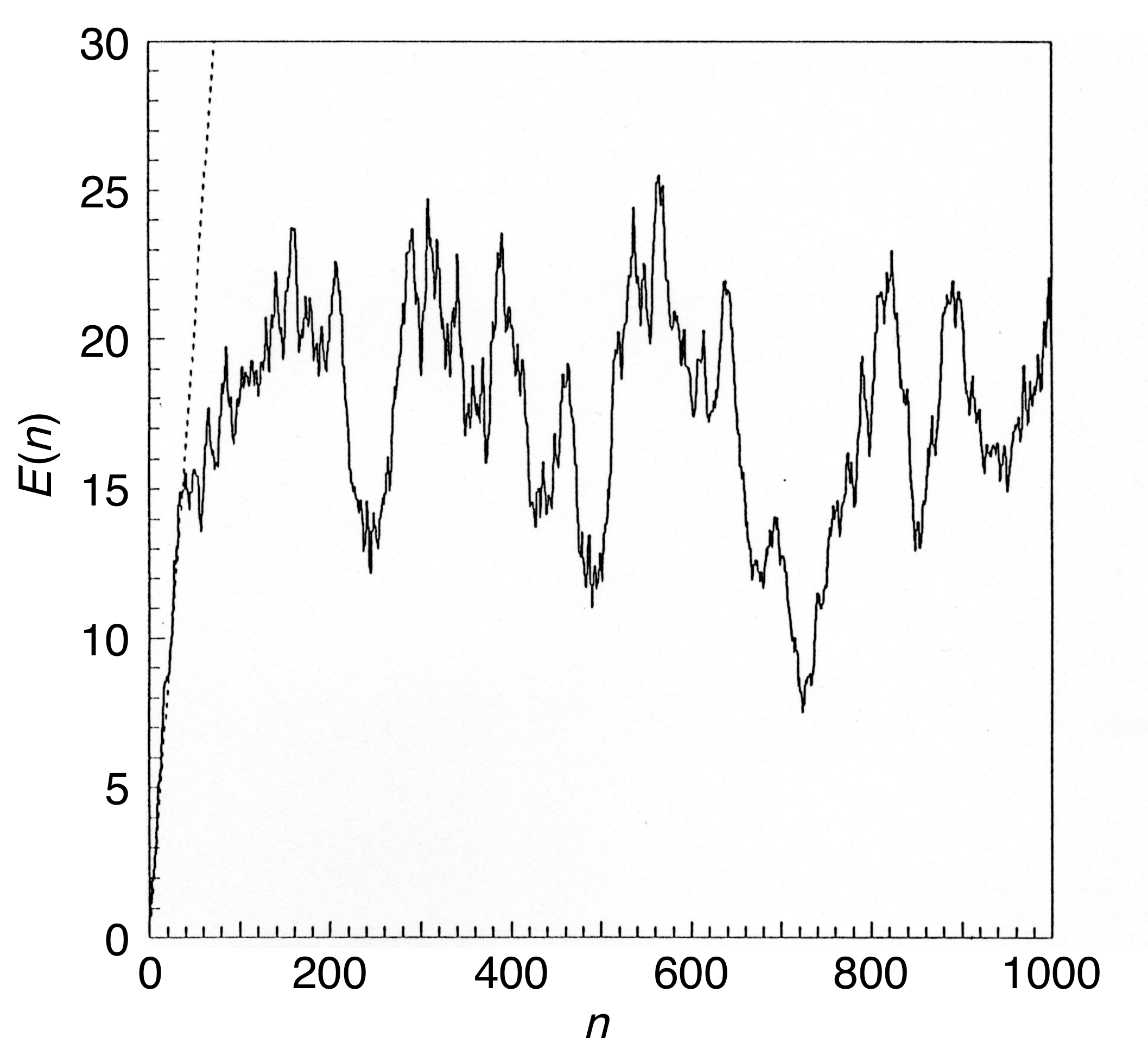}
\caption{Suppression of deterministic angular momentum diffusion in the quantum
kicked rotor. Time evolution of the mean kinetic energy, $E(n) = \langle p_n^2/2 \rangle$,
over the first 1000 time steps, for the classical kicked rotor, Eq.\ (\protect\ref{krhamil}),
(dotted) and its quantized version, Eq.\ (\protect\ref{rhomapwp}) (solid line).
The parameter values are $K = 10$ and $2\pi\hbar = 0.15/G$ ($G := (\sqrt{5} - 1)/2$).}
\label{figcqkrene}
\end{figure}

Operating on an infinite dimensional Hilbert space, the arguments explaining
quasi-periodicity of the time evolution generated by the quantum baker map do
not carry over immediately to the kicked rotor. On the contrary, one expects to see
a similar unbounded growth of the kinetic energy as symptom of chaotic diffusion as
in the classical standard map, in the regime of strong kicking. It was all the more surprising
for Casati \emph{et al.} \cite{CCIF83,She83} that their numerical experiments proved
the opposite: The linear increase of the kinetic energy ceases after a finite number of kicks
and gives way to an approximately steady state, with the kinetic energy fluctuating
in a quasi-periodic manner around a constantmean value (Fig.\ \protect\ref{figcqkrene}).

An explanation was found by analyzing the quasienergy eigenstates of the system
\cite{FGP82,FGP84,She86,CFGV86}. With a time-dependent external force,
the kicked rotor does not conserve energy. However, the invariance of the driving
under discrete translations of time, $t \to t+1$, allows to apply Floquet theory
\cite{Shi65,Zel66}. It implies the existence of eigenstates of $\hat U_{\rm QKR}$ with
unimodular eigenvalues $\exp({\rm i}\epsilon)$, that is, determined by
eigenphases $\epsilon$.

Quasienergy eigenstates can be calculated by numerical diagonalization of
$\hat U_{\rm QKR}$. It turns out that for generic values of the parameters, they
are exponentially localized: On average and superposed with strong fluctuations,
eigenstates $| \phi(\epsilon) \rangle$ have an exponential envelope of width $L$ around a
centre $l_{\rm c}(\epsilon)$,

\begin{equation} \label{qkdynloc}
\left\vert \langle l \vert \phi(\epsilon) \rangle\right\vert^2 \sim
\exp\left(- \frac{| l - l_{\rm c}(\epsilon) | }{L}\right).
\end{equation}

\noindent
The \emph{localization length} is approximately given by $L \approx (K/2\pi\hbar)^2$, hence
grows linearly with the classical diffusion constant, cf. Eq.\ \ref{krdiffconst}. Exponential
localization resembles Anderson localization, a phenomenon known from solid-state
physics \cite{And58,LR85}: In crystalline substances with sufficiently strong ``frozen
disorder'' (impurities, lattice dislocations, etc.), wavefunctions scattered at nonperiodic defects
superpose destructively, so that extended Bloch states compatible with the priodicity of
the lattice cannot build up. In the kicked rotor, however, the disorder required to prevent
extended states does not arise by any static randomness of a potential nor is it a consequence
of the chaotic dynamics of the classical map. It comes about by a dynamical effect
related to the nature of the sequence of phases $\hbar l^2 ({\rm mod} 2\pi)$ of the factor
$\hat U_{\rm rot}$ of the Floquet operator (\ref{qkrevop}). If the parameter $\hbar/2\pi$---
in the present context, it arises as a dimensionless quantity, Planck's constant in
units of a classical action---is not a rational, these phases constitute a pseudo-random
sequence. In one dimension, this disorder of number theoretical origin is strong enough
to localize eigenstates. Since the rationals form a dense subset of measure 0
of the real axis, an irrational value of $\hbar/2\pi$ is the generic case.

Even embedded in an infinite-dimensional Hilbert space, exponential localization
reduces the \emph{effective} Hilbert-space dimension to a finite number $D_\mathcal{H}$,
determined by the number of quasienergy eigenstates that overlap appreciably with a given
initial state. For a sharply localized initial state, say $|\psi(0)\rangle = | l_0 \rangle$,
it is given on average by $D_\mathcal{H} = 2L$. This explains immediately the crossover from
classical chaotic diffusion to localization described above: In the basis of localized eigenstates,
a sharp initial state overlaps with approximately $2L$ quasienergy states, resulting in
the same number of complex expansion coefficients. The initial ``conspiration of phases'',
required to construct the initial state $|\psi(0)\rangle = | l_0 \rangle$,
then disintegrates increasingly, with the envelope of the evolving state widening
diffusively, until all phases of the contributing eigenstates have lost their correlation
with the initial state, at a time $n^* \approx 2L$, in number of kicks.
The evolving state has then reached an exponential envelope,
similar to the shape of the eigenstates, Eq.\ (\ref{qkrevop}) (Fig.\ \protect\ref{figcmqkrpdis},
dashed lines), and its width fluctuates in a pseudo-random fashion, as implied by
the superposition of the $2L$ complex coefficients involved.

This scenario might appear as an exceptional effect, arising by the coincidence
of various special circumstances. Indeed, there exist a number of details and
exceptions, omitted in the present discussion, that lead to different dynamical
behaviour, such as accelerator modes in the classical model \cite{Kar83,IFZ02} and quantum
resonances for rational values of $\hbar/2\pi$ \cite{Izr90}. Notwithstanding, similar studies
of other models have accumulated overwhelming evidence that in quantum systems
evolving as a unitary dynamics, a permanent entropy production as in classical
chaos is excluded. In more abstract terms, this ``quantum death of classical chaos''
can be understood as the consequence of two fundamental principles:
the conservation of information under unitary time evolution, cf. App.\ \ref{qmentrocons},
a conservation law closely analogous to information conservation under
classical canonical transformations (App.\ \ref{clentrocons}), and the condition that 
the initial state contains only a finite amount of information.

This interpretation is corroborated by the global parameters characterizing
the behaviour of the quantum kicked rotor. In the presence of localization,
the dimension of the Hilbert space effectively accessible by an initial condition
local in angular momentum is $D_{\mathcal{H}} \approx 2L$. Even if the initial state
is a pure state with vanishing von-Neumann entropy, the maximum information
content it could achieve by incoherent processes or could produce by a quantum
dynamics imitating classical chaos is given by a homogeneous distribution over
$D_{\mathcal{H}}$ states, hence by $I(0) \approx c \ln(2L)$. Comparing this with
the entropy production by chaotic diffusion, Eq.\ (\ref{krentprod}), the cross-over
time $n^*$, in units of the kicking period, till the limited initial supply
of quantum entropy is exhausted, can be readily estimated. By equating

\begin{equation}
I(0) = I(n^*) = c\left[\ln\Bigl(\sqrt{2\pi D(K)n^*} / d_p \Bigr) + \frac{1}{2} \right],
\end{equation}

\noindent
 and setting $D(K) = K^2/2$, as in Eq.\ (\ref{krdiffconst}), and $d_p = \hbar$,
the angular momentum quantum, it turns out to be

\begin{equation} \label{crossovertinfo}
n^* \approx \frac{4}{\pi{\rm e}} K^2. 
\end{equation}

\noindent
It coincides exactly, as to the dependence on $K$, with similar estimates
based, e.g., on the energy-time uncertainty relation, and with numerical data,
which give

\begin{equation} \label{crossovertunctrty}
n^* \approx 2L \approx \frac{K^2}{2\pi^2\hbar^2}, 
\end{equation}

\noindent
and in order of magnitude even as to the prefactor.

\subsection{Breaking the splendid isolation: quantum chaos and quantum measurement}
\label{sec32}

If the absence of permanent entropy production in closed quantum systems
is interpreted as a manifestation of quantum coherence, it is natural to inquire
how immune this effect is to incoherent processes. They occur in a huge
variety of circumstances: in quantum systems embedded in a material environment,
as in molecular and solid state physics, interacting with a radiation field,
as in quantum optics, in dissipative quantum systems where decoherence
accompanies an irreversible energy loss, and most notably in all instances
of observation, be it by measurement in a laboratory or by leaving any kind of
permanent record in the environment \cite{Zur04}, even in the absence of a human observer.

In the present context, measurements are of particular interest, since they
allow to separate neatly two distinct phenomena, the loss of energy to the environment
and the exchange of entropy with it. Quantum measurement has been in the focus of
quantum theory from the early pioneering years on. It provides the indispensable
interface with the macroscopic world. The crucial step from quantum superpositions
to alternative classical facts remained an enigma for decades. The Copenhagen
interpretation includes the ``collapse of the wavepacket'' as an essential element
\cite{Boh28}, but treats it as an unquestionable postulate. The first systematic analysis
of quantum measurement by von Neumann \cite{Neu18} already provides a quantitative
description in terms of the density operator, rendering the wavepacket collapse
explicit as a reduction of the density matrix to its diagonal elements, but does not
yet illuminate the physical nature of this step, manifestly incompatible with the Schr\"odinger
equation. It was the contribution of Zurek and others \cite{Zur81,Zur82,Zur83,Zur84,HW87}
to interpret this process, in the spirit of quantum dissipation, as the consequence of the
interaction with the macroscopic number of degrees of freedom of the measurement apparatus
(the ``meter'') and its environment, to be described in a microscopic model as a heat bath
or reservoir. As one of the major implications of this picture, the collapse of the wavepacket
no longer appears as an unstructured point-like event but as a continuous process
that can be resolved in time \cite{Zur84}.

\subsubsection{Modelling continuous measurements on the quantum kicked rotor}
\label{sec321}

In this subsection, basic elements of this scheme will be adopted and applied to
the quantum kicked rotor in order to demonstrate how observation can thaw
dynamical localization and thus restore, at least partially, an entropy production
as in classical chaos. Reducing quantum measurement to the essential,
a continuous observation of the kicked rotor will be assumed, which leads to an
irreversible record of a suitable observable \cite{SS88}. Following established models of
quantum measurement \cite{HW87,UZ89,Zur81,Zur82,Zur83,Zur84}, these features can
be incorporated in a system-meter interaction Hamiltonian \cite{DG90B,DG90C,DG92}

\begin{equation} \label{sysmetintham}
H_{\rm SM} = g\, \hat x_{\rm M}\, \hat x_{\rm S}\, \Theta(t),
\end{equation}

\noindent
where $g$ controls the coupling strength and the Heaviside function $\Theta(t)$ switches
the measurement on at $t = 0$. The operator $\hat x_{\rm M}$, acting on the Hilbert
space of the meter, is the observable that indicates the measurement result
(its ``pointer operator'' \cite{Zur81,Zur82,Zur83,Zur84}), and $\hat x_{\rm S}$ is the measured
observable. In accord with the objective to study the impact of observation on localization
in angular momentum space, we shall focus on measurements of the angular
momentum $\hat l$. If the expectation $\langle l \rangle$ is observed as a
global measure,  this amounts to defining the measured operator as

\begin{equation} \label{sysmetinthampavg}
\hat x_{\rm S} = \hat l = \sum_{l=-\infty}^\infty l \vert l \rangle \langle l \vert.
\end{equation}

\noindent
Alternatively, a simultaneous observation of the full angular-momentum distribution
$P(l)$, so that the measurement affects homogeneously the entire
angular momentum axis, requires assuming a separate meter component
$\hat x_{{\rm M},l}$ for every eigenvalue of the angular momentum,

\begin{equation} \label{sysmetinthampdis}
H_{\rm SM} = g\, \hat {\bf x}_{\rm M} \cdot \hat {\bf x}_{\rm S}\, \Theta(t) =
g \sum_{l=-\infty}^\infty \hat x_{{\rm M},l} \hat x_{{\rm S},l}, \quad
\hat x_{{\rm S},l} = \vert l \rangle \langle l \vert.
\end{equation}

Some models of quantum measurement distinguish explicitly between the meter
proper, as a microscopic system interacting directly with the observed object, and a macroscopic
apparatus that couples in turn to the meter  \cite{HW87}, thus only indirectly to the object.
Such a distinction  is not necessary in the present context, it suffices to merge meter and
environment into a single macroscopic system. Moreover, we do not conceive a detailed
microscopic model of the meter as a heat bath (but see Sections \ref{sec42}, \ref{sec43} below),
starting instead directly from an evolution equation that takes the essential
consequences of the meter's macroscopic nature into account.

Specifically, the response of the meter is assumed to be Markovian,  that is,
to be immediate on the time-scales of the measured system, which in turn requires
the spectrum of the underlying heat bath to be sufficiently smooth. In terms
of the autocorrelation function of the meter operator $\hat x_{\rm M}$ \cite{DG90C},
that means

\begin{equation} \label{pointerautocorr}
\langle \hat x_{\rm M}(t) \hat x_{\rm M}(t') \rangle =
2T_{\rm M} \bigl\langle \hat x_{\rm M}^2 \bigr\rangle_0 \delta(t'-t),
\end{equation}

\noindent
denoting the autocorrelation time of $\hat x_{\rm M}$ as $T_{\rm M}$ and
the variance of its fluctuations in the uncoupled meter as
$\bigl\langle \hat x_{\rm M}^2 \bigr\rangle_0$. For the object,
coupled to the meter via Eqs.\ (\ref{sysmetinthampdis}) or (\ref{sysmetinthampavg}),
this already entails an irreversible dynamics. It can be represented as the time evolution
of the reduced density operator $\hat\rho_{\rm S}(t) = \Tr_{\rm M}\bigl(\hat\rho(t)\bigr)$.
In the interaction picture, $\hat\rho_{\rm S,I}(t) = \exp({\rm i}H_{\rm S}t/\hbar) \hat\rho_{\rm S}(t)
\exp(-{\rm i}H_{\rm S}t/\hbar)$ (transforming to a reference frame that follows
the proper dynamics generated by  $H_{\rm S}$, the Hamiltonian of the object),
Eq.\ (\ref{pointerautocorr}) implies a master equation of Lindblad type
\cite{DG90B,DG90C,DG92}

\begin{equation} \label{lindbladx}
\dot{\hat\rho}_{\rm S,I} = \gamma
\bigl[\hat x_{\rm S,I},[\hat\rho_{\rm S,I}, \hat x_{\rm S,I}]\bigr].
\end{equation}

\noindent
The parameter $\gamma = g^2 T_{\rm M} \langle \hat x_{\rm M}^2 \rangle_0$
has the meaning of a diffusion constant, as becomes evident by rewriting
Eq.\ (\ref{lindbladx}) in the representation of the operator canonically conjugate
to $\hat x_{\rm S} = \hat l_{\rm S}$, that is, of $\hat\theta$,

\begin{equation} \label{lindbladp}
\dot{\rho}_{\rm S,I}(\theta,t) =
\frac{\partial}{\partial  t}\, \langle \theta |\hat\rho_{\rm S}(t)| \theta \rangle =
\gamma\, \frac{\partial^2}{\partial \theta^2}\, \rho_{\rm S,I}(\theta,t).
\end{equation}

\noindent
The full master equation for the object density operator is then \cite{DG90B,DG90C,DG92}

\begin{equation} \label{master}
\dot{\hat\rho}_{\rm S} = -\frac{{\rm i}}{\hbar} [\hat H_{\rm S},\hat\rho_{\rm S}] +
\gamma \bigl[\hat x_{\rm S},[{\hat\rho}_{\rm S}, \hat x_{\rm S}]\bigr],
\end{equation}

\noindent
now including the unitary time evolution induced by $\hat H_{\rm S}$ through
the term $(-{\rm i} / \hbar) [\hat H_{\rm S},\hat\rho_{\rm S}]$. A quantum map
for the reduced density operator $\hat\rho_{\rm S}$ is obtained by integrating
the master equation over a single period of the driving. For the rotation phase
of the time evolution, between two subsequent kicks, Eq.\ (\ref{master}) yields
in the angular-momentum representation, for the case of a global angular-momentum
measurement, Eq.\ (\ref{sysmetinthampavg}),

\begin{equation} \label{rhomaprot}
\langle l' \vert \hat\rho'_{\rm S} \vert m' \rangle =
\exp\left(-\frac{{\rm i}\hbar}{2}(l'^2 - m'^2) - \gamma (l' - m')^2 \right)
\langle l' \vert \hat\rho_{\rm S} \vert m' \rangle
\end{equation}

\noindent
that is, off-diagonal matrix elements (often referred to as ``quantum coherences'')
decay with a rate determined by their distance $l-m$ from the diagonal and
the effective coupling $\gamma$. If the full distribution is measured,
see Eq.\ (\ref{sysmetinthampdis}), this step takes the form

\begin{equation} \label{rhomawprot}
\langle l' \vert \hat\rho'_{\rm S} \vert m' \rangle = \begin{cases}
\exp\left(-\frac{{\rm i}\hbar}{2}(l'^2 - m'^2) - \gamma\right)
\langle l' \vert \hat\rho_{\rm S} \vert m' \rangle &\text{$l' \neq m'$}, \\
\langle l' \vert \hat\rho_{\rm S} \vert l' \rangle &\text{$l' = m'$}. \end{cases}
\end{equation}

\begin{figure}[H]
\centering
\includegraphics[width=15 cm]{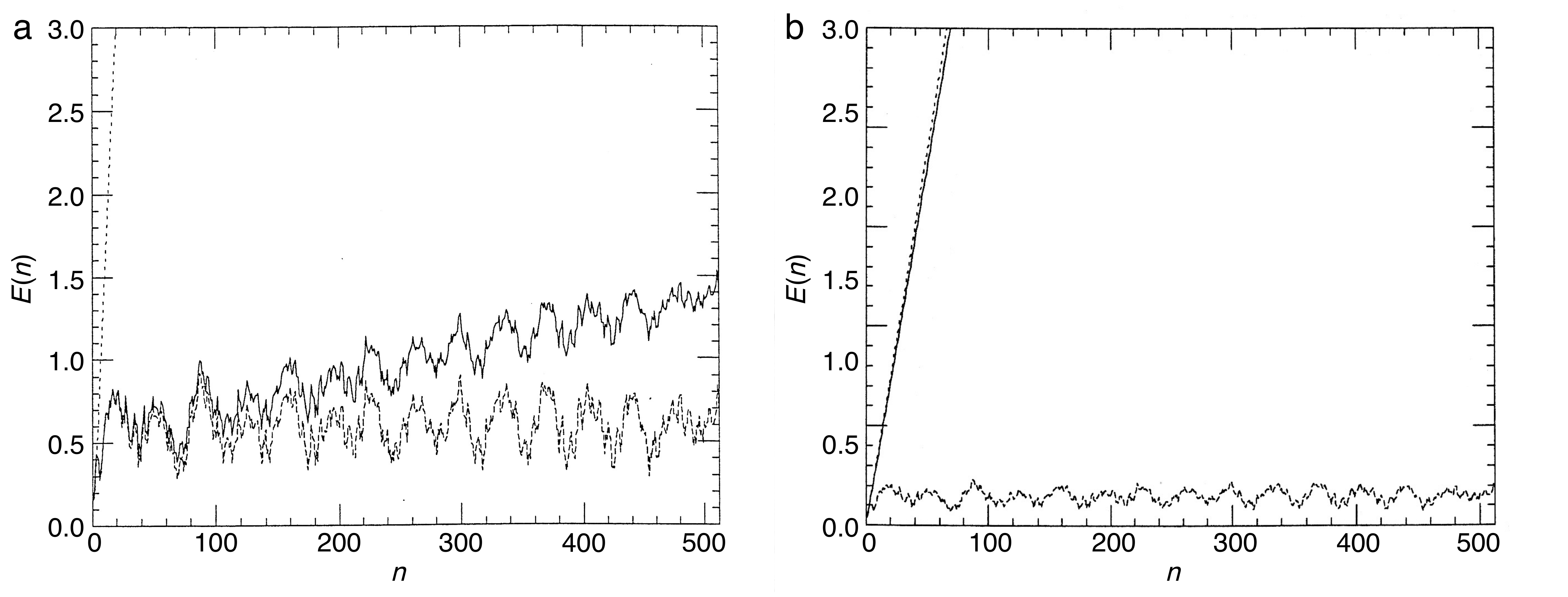}
\caption{Deterministic angular momentum diffusion is revived in the quantum
kicked rotor with continuous measurements. Time evolution of the mean kinetic energy,
$E(n) = \langle p_n^2/2 \rangle$, over the first 512 time steps for the measured
dynamics of the quantum kicked rotor, Eq.\ (\protect\ref{rhomapwp}) (solid line), the
stochastic classical map, Eqs.\ (\protect\ref{noisystandardmap},\protect\ref{noisevariancewp})
(dotted line), and the unobserved dynamics of the quantum kicked rotor,
Eq.\ (\protect\ref{rhomapwp}) (dashed line), for (\textbf{a}) weak
vs.\ (\textbf{b}) strong effective coupling. A continuous measurement of
the full action distribution was assumed. The parameter values are
$K = 5$, $2\pi\hbar = 0.1/G$ ($G := (\sqrt{5} - 1)/2$), and $\nu = 10^{-3}$ (a),
$\nu = 0.5$ (b).}
\label{figcmqkrene}
\end{figure}   

\noindent
The kicks are too short to be affected by decoherence, their effect on the
evolution of the density matrix results from the unitary term in Eq.\ (\ref{master})
alone. The integration over the $\theta$-dependent kicks is conveniently performed
by switching from the $l$- to the $\theta$-representation and back again,
resulting in

\begin{equation} \label{rhomapkick}
\langle l'' \vert \hat\rho''_{\rm S} \vert m''\rangle =
\sum_{l',m' = -\infty}^\infty b_{l''-l'}(k) b^*_{m''-m'}(k)
\langle l' \vert \hat\rho'_{\rm S} \vert m' \rangle.
\end{equation}

\noindent
The Bessel functions $b_n(x) = {\rm i}^n J_n(x)$ from the integration over $\theta$.
The full quantum map is obtained concatenating Eqs.\ (\ref{rhomaprot}) or (\ref{rhomawprot})
with (\ref{rhomapkick}). For measurements of $\langle l \rangle$, it reads

\begin{equation} \label{rhomapp}
\langle l \vert \hat\rho_{{\rm S},n+1} \vert m\rangle =
\sum_{l',m' = -\infty}^\infty b_{l'-l}(k) b^*_{m'-m}(k)
\exp\left(-\frac{{\rm i}\hbar}{2}(l'^2 - m'^2) - \gamma (l' - m')^2 \right)
\langle l' \vert \hat\rho_{{\rm S},n} \vert l' \rangle,
\end{equation}

\noindent
while for measurements of $P(l)$,

\begin{align} \label{rhomapwp}
\langle l \vert \hat\rho_{{\rm S},n+1} \vert m\rangle =
\sum_{l',m' = -\infty}^\infty & b_{l'-l}(k) b^*_{m'-m}(k) \nonumber \\
& \left[\exp\left(-\frac{{\rm i}\hbar}{2}(l'^2 - m'^2) \right) - e^{-\gamma} (1 - \delta_{m'-l'})
+ \delta_{m'-l'}\right] \langle l' \vert \hat\rho_{{\rm S},n} \vert m' \rangle.
\end{align}

\begin{figure}[H]
\centering
\includegraphics[width=15 cm]{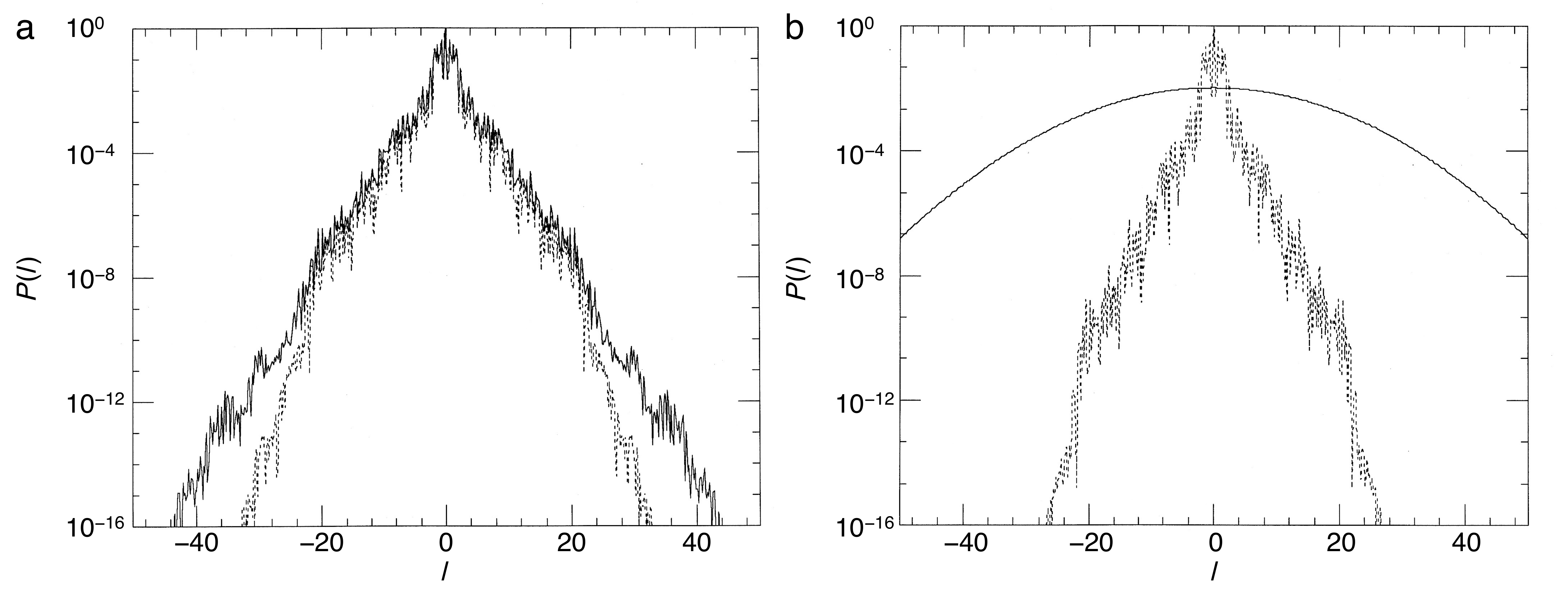}
\caption{Dynamical localization is destroyed in the quantum kicked rotor with continuous
measurements. Probability distribution $P(l)$ of the angular momentum $l$,
after the first 512 time steps, for the measured dynamics of the quantum kicked rotor,
Eq.\ (\protect\ref{rhomapwp}) (solid lines), compared to the unmeasured dynamics of
the same system, Eq.\ (\protect\ref{rhomapwp}) (dashed), for (\textbf{a}) weak
vs.\ (\textbf{b}) strong effective coupling. A continuous measurement of
the full action distribution was assumed. The parameter values are
$K = 5$, $2\pi\hbar = 0.1/G$ ($G := (\sqrt{5} - 1)/2$), and $\nu = 10^{-4}$ (a),
$\nu = 0.5$ (b).}
\label{figcmqkrpdis}
\end{figure}   

The map alternates the unitary time evolution of the quantum kicked rotor with
incoherent steps that lead to a gradual decay of the non-diagonal elements
of the density matrix. In the limit of strong effective coupling to the meter,
$\gamma \gg 1$, corresponding to a high-accuracy measurement of the angular
momentum, the density matrix is completely diagonalized anew at each time step,
and the object system leaves the measurement in an incoherent superposition
of angular-momentum states, as required by the principles of quantum measurement
(Figs.\ \protect\ref{figcmqkrene}b, \protect\ref{figcmqkrpdis}b).
For a weaker coupling, the loss of coherence per step is only partial, restricting
the density matrix to a diagonal band with a Gaussian profile of width
$\sim \gamma^{-1}$, if $\langle l \rangle$ is measured, or reducing its off-diagonal
elements homogeneously by $e^{-\gamma}$, if the full distribution is recorded
(Figs.\ \protect\ref{figcmqkrene}a, \protect\ref{figcmqkrpdis}a).
In any case, decoherence in the angular momentum representation is equivalent
to a diffusive spreading of the angle $\theta$. It imitates the action of classical
chaos in that it effectively destroys the autocorrelation of the angle variable.

The framework set by Eq.\ (\ref{master}) is easily extended to include dissipation
\cite{DG86,DG87,DG90A}. An additional term, proportional to the friction constant $\lambda$,

\begin{align} \label{dissmaster}
\dot{\hat\rho}_{\rm S} = &-\frac{{\rm i}}{\hbar} [\hat H_{\rm S},\hat\rho_{\rm S}] +
\gamma \bigl[\hat x_{\rm S},[{\hat\rho}_{\rm S}, \hat x_{\rm S}]\bigr] \nonumber\\
&+ \frac{1}{2} g^2 \lambda
\left(\bigl[\hat x_{\rm S} \hat\rho_{\rm S},[\hat H_{\rm S},\hat x_{\rm S}]\bigr] -
\bigl[[H_{\rm S},\hat x_{\rm S}],\hat\rho_{\rm S}\hat x_{\rm S}\bigr]\right),
\end{align}

\noindent
induces incoherent transitions between angular momentum eigenstates
towards lower values of $l$, modelling Ohmic friction with a damping constant
$\lambda$, as in the classical standard map with dissipation,
Eqs.\ (\ref{disstandardmap},\ref{disstandardnewton},\ref{krrhofopl}) \cite{Zas78,SW85}.
In terms of a classical stochastic dynamics, to be detailed in
the following subsection, it corresponds to a drift of the probability density
in phase space towards lower angular momentum.

\subsubsection{Semiclassical Langevin approximation for the measured quantum dynamics}
\label{sec322}

Describing the quantum dynamics in terms of a master equation for
the reduced density operator only provides a global statistical account. However,
in the semiclassical regime of small $\hbar$, compared to the periodicity in $p$ of
the classical phase space, it can be replaced by an approximate description as a classical
Langevin equation with a noise term of quantum origin that induces diffusion in $\theta$ \cite{DG90B,DG90C,DG92}. In this limit, the Wigner function, which represents the density
operator in a quantum equivalent of classical phase space (with quantized momentum, though),
evolves as a phase-space flow following classical trajectories, as does the corresponding
classical phase-space density, but superposed with a random quivering. These trajectories
are adequately described by a noisy standard map similar to Eq.\ (\ref{standardmap})
\cite{DG90B,DG90C,DG92},

\begin{equation} \label{noisystandardmap}
\begin{pmatrix}p_{n+1} \\ \theta_{n+1} \end{pmatrix} = 
\begin{pmatrix}p_n + K \sin(\theta_{n+1}) \\ \theta_n + p_n + \xi_n \end{pmatrix},
\end{equation}

\noindent
now including a random process $\xi_n$ with mean $\langle \xi_n \rangle = 0$,
distributed as a Gaussian with variance $\langle \xi_n \xi_n' \rangle =
\hbar^2\gamma\delta_{n'-n}$ for measurements of $\langle p\rangle$, or 

\begin{equation} \label{noisevariancewp}
\xi_n = \begin{cases}
0 &\text{with probability $\nu$,} \\
\text{equidistributed in $[0,1[$} & \text{with probability $1-\nu$,}
\end{cases}
\end{equation}

\noindent
with $\nu = 1 - e^{-\gamma}$, if $P(l)$ is measured. If Ohmic friction is taken
into account, as in the master equation (\ref{dissmaster}),
the noisy map (\ref{noisystandardmap}) acquires a damping of
the angular momentum per time step by a factor $\exp(-\lambda)$,

\begin{equation} \label{dissnoisystandardmap}
\begin{pmatrix}p_{n+1} \\ \theta_{n+1} \end{pmatrix} = 
\begin{pmatrix}p_n + K \sin(\theta_{n+1}) \\ \theta_n + e^{-\lambda}p_n + \xi_n \end{pmatrix}.
\end{equation}

\subsubsection{Numerical results}\label{sec323}

\begin{figure}[H]
\centering
\includegraphics[width=15 cm]{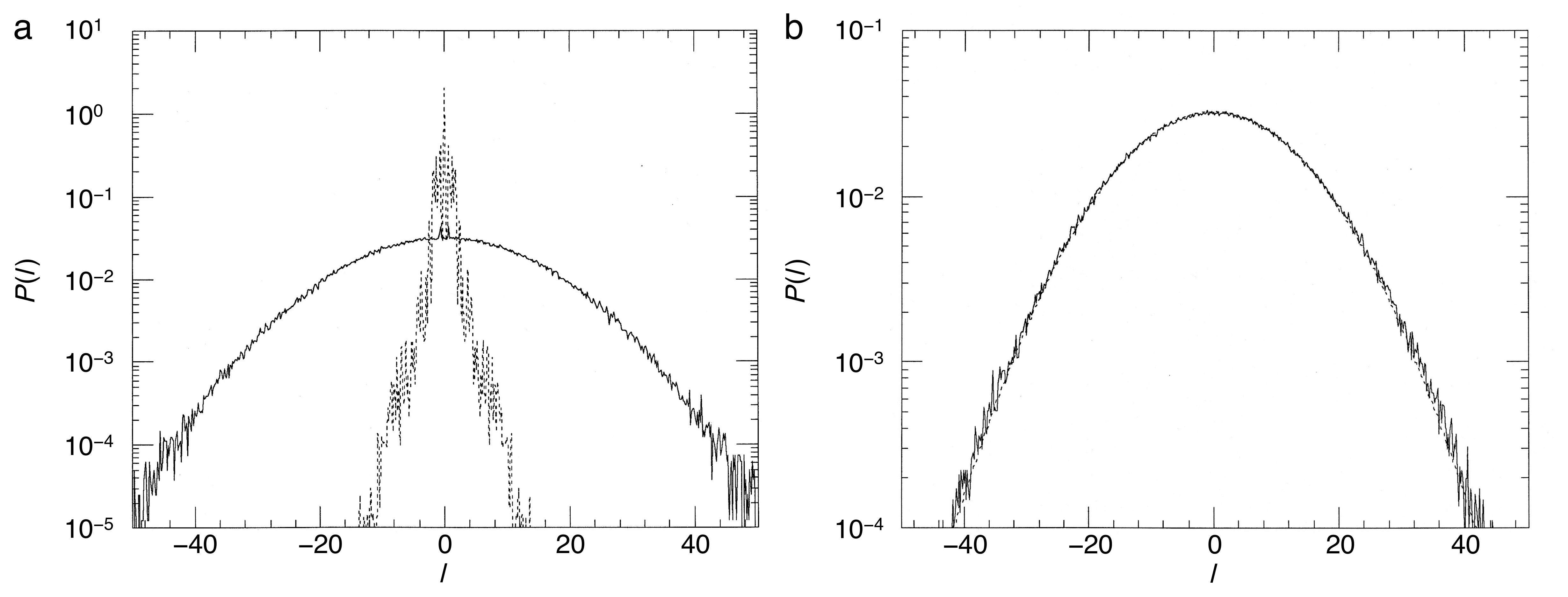}
\caption{Probability distribution $P(l)$ of the momentum $l$, after the first 512
time steps, for the measured dynamics of the quantum kicked rotor,
Eq.\ (\protect\ref{rhomapwp}) (dotted lines), compared to the stochastic classical map,
Eqs.\ (\protect\ref{noisystandardmap},\protect\ref{noisevariancewp})
(solid lines), for (\textbf{a}) weak vs.\ (\textbf{b}) strong effective coupling.
A continuous measurement of the full action distribution was assumed.
The parameter values are $K = 10$, $2\pi\hbar = 0.1/G$ ($G := (\sqrt{5} - 1)/2$),
and $\nu = 10^{-4}$ (a), $\nu = 0.5$ (b).}
\label{figscmqkrpdis}
\end{figure}

Numerical experiments performed with both, the quantum map for the density
matrix, Eqs.\ (\ref{rhomapp},\ref{rhomapwp}), and its semiclassical approximation,
Eqs.\ (\ref{noisystandardmap},\ref{dissnoisystandardmap}), give a detailed picture of
the effect of continuous observation on quantum chaos \cite{DG90B,DG90C,DG92}. Figure \protect\ref{figcmqkrene} compares the time dependence of the mean kinetic energy for
the quantum kicked rotor, Eq.\ (\ref{qkrevop}) (dashed lines), the same system under
continuous measurement, Eq.\ (\ref{rhomapwp}) (solid lines), and the stochastic
classical map, Eqs.\ (\protect\ref{noisystandardmap},\protect\ref{noisevariancewp}) (dotted).
Above all, the data shown provide clear evidence that \emph{incoherent processes
induced by measurements destroy dynamical localization}. Even for weak coupling
to the apparatus, Figs.\ \protect\ref{figcmqkrene}a, \protect\ref{figcmqkrpdis}a, classical
angular momentum diffusion is recovered, albeit on a time scale
$n_{\rm c} \approx \nu^{-1}$, much larger than the cross-over time $n^*$,
cf.\ Eq.\ (\ref{crossovertunctrty}), if $\nu \ll 1/2L$, and with a diffusion constant
$D_{\rm qm} \approx D(K) n^*/n_{\rm c}$, reduced accordingly with respect
to its classical value $D(K)$. For stronger coupling, the
measurement-induced diffusion approaches the classical strength $D(K)$.
Since it randomizes the angle variable indiscriminately, erasing all fine structure
in classical phase space, it ignores deviations of $D(K)$ from the gross
estimate (\ref{krdiffconst}), caused, e.g., by accelerator modes of the classical
standard map \cite{Kar83,IFZ02}. In fact, measurement-induced diffusion occurs already
for kick strengths $K < K_{\rm c}$, below the classical threshold to chaotic diffusion
$K_{\rm c} \approx 1$, where in the exact classical map, diffusion is still blocked
by regular tori extending across the full range $\theta \in [0,2\pi[$.
Moreover, Fig.\ \protect\ref{figscmqkrpdis}b, showing the angular momentum
distribution after 512 time steps, demonstrates that at this stage, the typical
$\exp(-|l|/L)$ shape indicating localization has given way to a Gaussian
envelope, characteristic of diffusion.

Figure \ref{figscmqkrpdis} compares the angular momentum reached after 512 time steps
for the measured quantum system in the description by the master equation
(\ref{rhomapwp}) (dotted lines) with that obtained for the noisy map (\ref{noisystandardmap})
(solid lines). For sufficiently strong coupling, Fig.\ \protect\ref{figscmqkrpdis}b, it is faithfully
reproduced by the semiclassical Langevin equation (\ref{noisystandardmap}),
as is the overall energy growth, see Fig.\ \protect\ref{figcmqkrene}b (dotted line).

The diffusion constant of the measurement-induced angular momentum diffusion
also allows us to estimate directly the entropy produced by the measured
quantum system: Replacing in Eq.\ (\ref{krentprod}) the classical diffusion constant
$D(K)$ by the reduced quantum mechanical value $D_{\rm qm}$ yields

\begin{equation} \label{qkrmentprod}
I(t) = \frac{c}{2}\left[\ln\left(\frac{2\pi D_{\rm qm} t}{d_p^2}\right) +
2\ln\left(\frac{n^*}{n_{\rm c}}\right) + 1\right],
\end{equation}

\noindent
As the production rate for diffusive spreading is independent of the diffusion constant,
it is here the same as for the classical standard map, $\dot I(t) = c/2t$.
Such a positive entropy production is not compatible with entropy conservation in closed
quantum systems, App.\ \ref{qmentrocons}. The only possible explanation therefore
refers to the measured quantum system \emph{not} being closed, so that the entropy
generated actually infiltrates from the macroscopic meter to which it is coupled.
This interpretation becomes plausible also considering the fact that obviously,
there must be an entropy flow from the object towards the meter---or else the
measured data could not reach it: There is no reason why the information current 
from object to meter should not be accompanied by an opposite current, from
meter to object.

The three phases of the time evolution of, in particular, the weakly (i.e., with small
coupling to the meter) measured quantum kicked rotor can now be interpreted
from the point of view of entropy flows: During the initial phase, $n \lesssim n^*$,
the quantum map follows closely the classical standard map, producing entropy
from its own supply provided by the initial state. Once this supply is exhausted, at
$n \approx n^*$, entropy production stalls, the system localizes and crosses over
to quasi-periodic fluctuations. Only on a much longer time-scale, for
$n \gtrsim n_{\rm c} \gg n^*$, sufficient entropy infiltrates from the meter
to become manifest again in the dynamics of the kicked rotor as diffusive
angular momentum spreading.

\begin{figure}[H]
\centering
\includegraphics[width=15.5 cm]{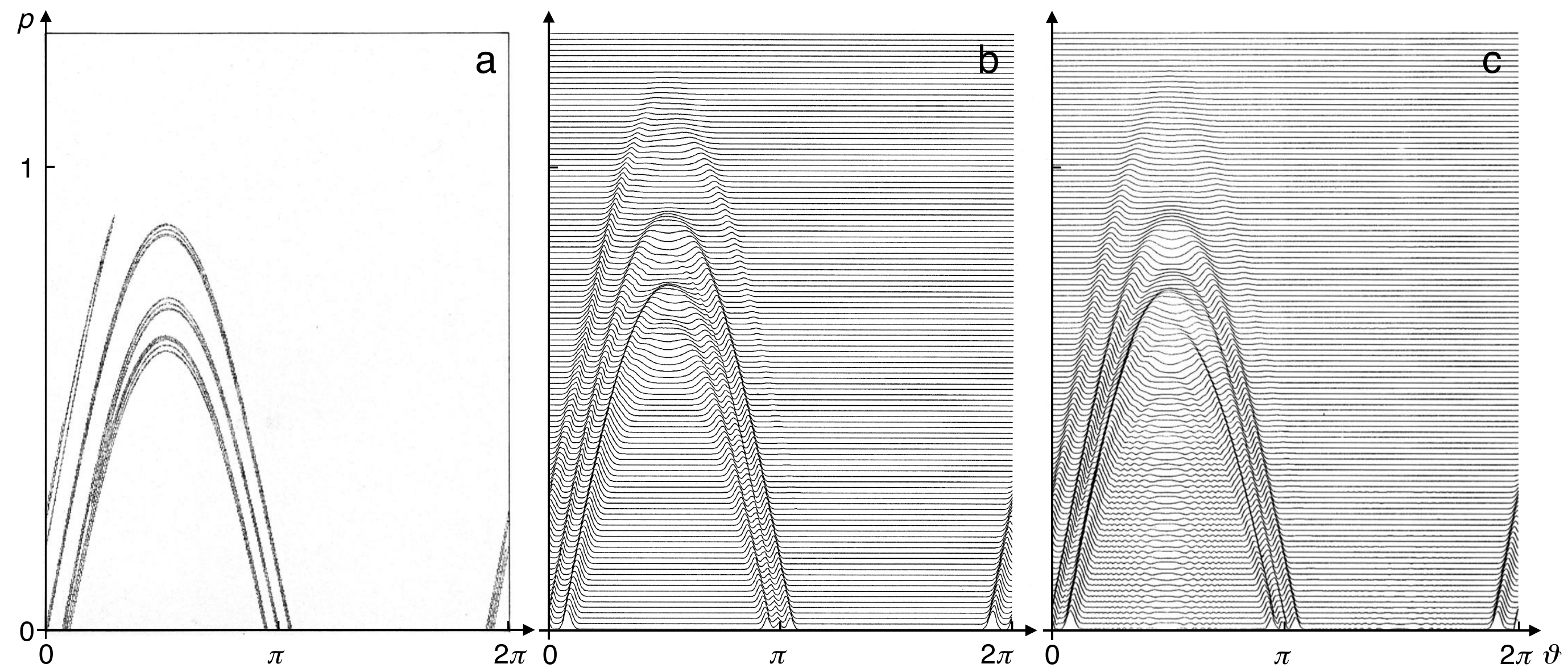}
\caption{Classical and quantum stationary-state distributions of the
dissipative standard map for $n \gg 1/\lambda$. (\textbf{a}) Support of
the strange attractor of the classical map (\protect\ref{disstandardmap})
in $(p,\theta)$ phase space. (\textbf{b}) Stationary state of
the semiclassical stochastic map (\protect\ref{dissnoisystandardmap}),
plotted at discrete angular momentum values $p_l = \hbar l$, as in panel (c).
(\textbf{c}) Long-time limit of the density operator for the master equation
(\protect\ref{dissmaster}), represented as the corresponding Wigner function,
which has support along the quantized angular momentum values
$l \hbar$, $l \in \mathbb{Z}$. The parameter values are $n = 10$, $K = 5$,
$\lambda = 0.3$, and $2\pi\hbar = 0.02$ (b,c). Only the upper (positive-momentum,
$p \geq 0$) part of phase space is shown, the lower ($p leq 0$) part is related to
it by parity, $p \to -p$, $\theta \to -\theta$.}
\label{figcsqkrstratt}
\end{figure}

Incorporating friction gives the opportunity to take a look also at the modifications
of classical dissipative chaos with that are required by quantization,
in particular of the fractal geometry of strange attractors. The master equation
(\ref{dissmaster}) as well as the stochastic semiclassical approximation,
Eq.\ (\ref{dissnoisystandardmap}), can be solved numerically and compared with
the classical dissipative standard map (\ref{disstandardmap}) \cite{DG86,DG87,DG90A}.
Fig.\ \protect\ref{figcsqkrstratt} compares the stationary states approached by
these maps for $n \gg 1/\lambda$, the time scale of contraction onto
the attractor. The classical strange attractor, Fig.\ \protect\ref{figcsqkrstratt}a,
here represented as its support in $(p,\theta)$ phase space, roughly follows
a ($-\sin\theta$)-curve. The stationary state of the full quantum master equation,
depicted as the Wigner function corresponding to the stationary density operator,
Fig.\ \protect\ref{figcsqkrstratt}c, shows a smoothed structure that eliminates
the self-similarity of the classical fractal geometry. The wavy modulations
visible in panel (c) are owed to the tendency of Wigner function to exhibit
fringes where it takes negative values, if the support of the positive regions is strongly
curved. They are absent in the stationary state of the semiclassical noisy map, panel (b).

\section{Quantum measurement and quantum randomness in a unitary setting}\label{sec4}

In the examples discussed in the preceding sections, the central issue was chaotic entropy
production and its suppression by coherence effects in closed quantum systems. Measurement
served as a particular case of interaction with a macroscopic environment, giving
rise to a two-way exchange of information. A transfer of information on the state of the object is
the essence of measurement. It does not even require a human observer, the physical
environment can play the r\^ole of the ``witness'' \cite{Zur04}. Conversely, entropy
entering the measured object from the side of the apparatus imparts a stochastic component
to the proper dynamics of the object \cite{UZ89}. Quantum chaos is specially sensitive to this
effect, as it amplifies even minuscule amounts of entropy penetrating from outside
and displays them directly as a drastic change of behaviour.

The present section takes up this idea to apply it within the context of quantum
measurement, to situations where inherent instabilities of the measurement process itself,
instead of a sensitive dependence on initial conditions of a measured chaotic system,
let us expect similar effects as in the case of quantum chaos. It is not obvious, though,
where in the context of measurement instabilities should exist, of a kind even
remotely comparable to chaotic dynamics. To see this, a final step has to be added
to the above outline of the quantum measurement process.

\subsection{Quantum randomness from quantum measurement}\label{sec41}

The collapse of the wavepacket is not only incompatible with a unitary time evolution,
it also violates the conservation of entropy (App.\ \ref{qmentrocons}).
If the measured system is initiated in a pure state, 

\begin{equation} \label{inimeasure}
\vert \psi_{\rm S,ini} \rangle = \sum_\alpha a_\alpha \vert \alpha \rangle,
\end{equation}

\noindent
(assuming a discrete basis of eigenstates of the measured operator,
$\hat x \vert \alpha \rangle = x_\alpha \vert \alpha \rangle$, $\alpha \in \mathbb{Z}$) a complete
collapse leads to a mixed state comprising the same components,

\begin{equation} \label{psicollapse}
\hat\rho_{\rm S,ini} = \vert \psi_{\rm S,ini}  \rangle \langle \psi_{\rm S,ini} \vert \to
\hat\rho_{\rm S,coll} = \sum_\alpha p_\alpha \vert \alpha \rangle
\langle \alpha \vert,\quad p_\alpha = |a_\alpha|^2.
\end{equation}

\noindent
The increase in entropy from the pure initial state ($I_{\rm ini} = 0$) is thus

\begin{equation} \label{collapseinfo}
I_{\rm coll} = -c\Tr\left(\hat\rho_{\rm S,coll} \ln(\hat\rho_{\rm S,coll})\right) =
                     -c\sum_\alpha p_\alpha \ln(p_\alpha).
\end{equation}

\noindent
It is readily explained and can be modelled in microscopic detail as a consequence
of the entanglement of the object with the macroscopic apparatus
\cite{HW87,UZ89,Zur81,Zur82,Zur83,Zur84}, which in
the reduced density operator of the object becomes manifest as information gain. The
density operator, reduced to its diagonal,
$\langle \alpha \vert \hat\rho_{\rm S,coll} \vert \alpha' \rangle =
p_\alpha \delta_{\alpha'-\alpha}$, 
is interpreted as a set of probabilities $p_\alpha$ for the measurement resulting in
the eigenvalue $x_\alpha$ of the measured operator $\hat x$.

With this step, the measurement is not yet complete. From the Copenhagen interpretation
onwards \cite{Boh28}, all quantum measurement schemes add a crucial final transition,
to the object exiting the process again in a pure state, one of the eigenstates
$\vert \alpha \rangle$,

\begin{equation} \label{secondcollapse}
\hat\rho_{\rm S,coll} = \sum_\alpha | p_\alpha \alpha \rangle \langle \alpha | \to \hat\rho_{\rm S,fin} =
\begin{cases}
\phantom{| \alpha \rangle}\vdots\phantom{\langle \alpha |} & \quad\quad\quad\quad\vdots \\
| \alpha \rangle \langle \alpha | & \text{with probability $p_\alpha$} \\
\phantom{| \alpha \rangle}\vdots\phantom{\langle \alpha |} & \quad\quad\quad\quad\vdots 
\end{cases}
\end{equation}

\noindent
returning the information content to its initial value, $I_{\rm fin} = I_{\rm ini} = 0$.
This step is sometimes referred to as ``second collapse of the wavepacket''.
In contrast to the ``first collapse'', though, it is usually considered to be of little interest
for the discussion of fundamentals of quantum mechanics, since it appears as a mere
classical random process, analogous to drawing from an urn.

Indeed, on the face of it, there is not even a credit left in the information balance
between initial and final states. Both are pure. However, the random process
behind the phrase ``with probability $p_\alpha$'' does have a quantum
mechanical side to it. This applies at least to all measurements of operators
with a discrete spectrum, such as, for example, the angular momentum $\hat l$
featured in the context of the kicked rotor.

It becomes particularly evident in the case of operators on finite-dimensional
Hilbert spaces, notably and as the simplest possible instance, two-state systems
(``qbits''), say $\mathcal{H} = {\rm span}\{|\!\!\downarrow\rangle,|\!\!\uparrow\rangle\}$,
$\hat x | \!\!\downarrow\rangle = -\frac{\hbar}{2} | \!\!\downarrow\rangle$, 
$\hat x | \!\!\uparrow\rangle = \frac{\hbar}{2} | \!\!\uparrow\rangle$. 
If the initial state is a Schr\"odinger cat, neutral with respect to measurements of $\hat x$,

\begin{equation} \label{inispin}
\vert \psi_{\rm S,ini} \rangle = \frac{1}{\sqrt{2}}
(| \!\!\downarrow\rangle \pm |\!\!\uparrow\rangle\rangle),
\end{equation}

\noindent
the results $|\!\!\downarrow \rangle \langle \downarrow\!\! |$ and
$|\!\!\uparrow \rangle \langle \uparrow\!\! |$ are expected with equal probabilities
$p_\downarrow = p_\uparrow = 0.5$. While each outcome is a pure state with
definite eigenvalue, repeated measurements of an ensemble of systems in
the same initial state result in a random binary sequence, distinguished as
``quantum randomness'' and considered unpredictable in a more fundamental
sense than any classical stochastic process \cite{BK&18}. The von-Neumann entropy,
as canonical measure of the information contained in a quantum system,
is not able to capture the difference between a pure state resulting from
a deterministic preparation and an element of a sequence of
pure states which, as an ensemble, represent a prototypical random process.

The mere existence of a set of privileged states, the eigenstates of the measured operator
(forming the ``pointer basis'', a term coined by Zurek \cite{Zur81,Zur82,Zur83,Zur84}),
of course does not imply any instability. To be sure, the conservation under
unitary transformations of the overlap $\langle\phi | \psi\rangle$ as a measure of distance
between two states $|\ psi\rangle$, $| \phi\rangle$ ensures that
there cannot be any attractors or repellers in Hilbert space \cite{Per95}. This situation changes,
however, as soon as the non-unitary dynamics of incoherent processes in the
projective Hilbert space is concerned. In quantum measurement, in particular,
the \emph{quantum Zeno effect} \cite{MS77,IHBW90} plays a pivotal r\^ole \cite{Zur82}: If 
a measurement is made on a state vector that is about to rotate away from a pointer-basis
state it has been prepared in, for example by a previous measurement, this subsequent
measurement will project the state back to the nearest pointer basis state as indicated by
Eq.\ (\ref{secondcollapse}) \cite{Zur81,Zur82,Zur83,Zur84}, that is, the state it just
departed from. The more frequently the same measurement is being repeated,
the stronger will be its stabilizing effect towards the initial pointer state:
it thus becomes an attractor in the projective Hilbert space of
the measured object \cite{Zur81,Zur82}.

If there is not just a single such state but a finite or even countably infinite
number of attractors, it is clear that their basins of attraction in projective Hilbert space
must be separated by boundaries, manifolds along which the system is unstable.
For example, for a two-state system, the projective Hilbert space is the Bloch sphere,
its poles representing the pointer states, hence the attractors for measurements of
the vertical spin component (Fig.\ \protect\ref{figbloch}). Symmetry already implies that
the boundary separating their basins of attraction, the two hemispheres, must be the equator,
representing the manifold all Schr\"odinger-cat states as defined in Eq.\ (\ref{inispin}).
Of course, the attraction towards the poles is strongest in their immediate neighbourhood and
vanishes for states orthogonal to the pointer states, as applies to all states along the equator.

The description in terms of an evolution equation for the density operator,
such as the master equation (\ref{master}), however does not allow to go beyond stating
likelihoods, in this example equal probabilities for the two outcomes. Otherwise, it leaves
the second collapse as a black box. A more profound analysis is possible, though, by going
to a detailed microscopic account of the coupled object-meter system. Since
this comprehensive system is closed as a whole, it not only permits a description in the
framework of unitary time evolution. The conservation of entropy moreover opens
the possibility to follow the information interchanged between the two subsystems.

\subsection{Spin measurement in a unitary setting}\label{sec42}

The setup sketched in Sect.\ \ref{sec321} is a suitable starting point for a
model of measurements on a two-state system. In order to include a microscopic
account of the meter, it is broken down into a set of, say, harmonic oscillators
with frequencies $\omega_n$. The measurement object now reduces
to a spin-$\frac{1}{2}$ system. Modifying the object-meter coupling,
Eqs.\ (\ref{sysmetinthampdis},\ref{sysmetinthampavg}) accordingly, it now takes the form

\begin{equation} \label{spinmetintham}
H_{\rm SM} = \sum_n  g_n \hat\sigma_z (\hat a_n^\dagger + \hat a_n)\, \Theta(t),
\end{equation}

\noindent
where the measured observable is specified as $\hat x_{\rm S} = \hat\sigma_z$,
the vertical spin component, coupled with a strength $g_n$ to meter operators
$\hat x_{{\rm M},n} = \hat a_n^\dagger + \hat a_n$ (the position operators of
the $n$th mode of the meter, up to a factor $\sqrt{2}$). Complemented by self-energies
$H_{\rm S} = \frac{1}{2} \hbar\omega_0\hat\sigma_x$ of the object and
$H_{\rm M} = \sum_n \hbar\omega_n \left(\hat a_n^\dagger \hat a_n +
\frac{1}{2}\right)$ of the meter, a total Hamiltonian for the measurement process

\begin{align} \label{spinmetham}
H &= H_{\rm S} + H_{\rm SM} + H_{\rm M} \nonumber\\
&=  \frac{1}{2} \hbar\omega_0\hat\sigma_x +
\sum_n  g_n \hat\sigma_z (\hat a_n^\dagger + \hat a_n)\, \Theta(t) +
\sum_n \hbar\omega_n \left(\hat a_n^\dagger \hat a_n + \frac{1}{2}\right)
\end{align}

\noindent
results. In terms of quantum optics, for instance, it can be interpreted as describing a two-level
atom interacting with a microwave cavity supporting discrete modes $n$ \cite{RBH97}.

The model is not complete without specifying the initial state of the total system.
Supposing that it factorizes between object and meter \cite{Neu18,Zur81,Zur82,HW87}, 

\begin{equation} \label{spinmetini}
|\Psi_{\rm ini}\rangle = |\psi_{\rm S,ini}\rangle |\psi_{\rm M,ini}\rangle,
\end{equation}

\noindent
the initial states of the two components can be defined separately. 
For the object, assume a state that is neutral with respect to measurements
of $\hat\sigma_z$, as in Eq.\ (\ref{inispin}). The initial state of the meter
should not introduce a spatial bias of position or momentum, either, so that
$\langle\hat x_{\rm M}\rangle = 0$, $\langle\hat p_{\rm M}\rangle = 0$,
but otherwise can be an arbitrary coherent superposition of harmonic oscillator states.

A crucial issue concerning Hamiltonian and the initial condition is their symmetry under
spatial reflections with respect to the direction of the vertical spin component, $z \to -z$.
The total Hamiltonian as well as the initial state of the object should be invariant under
this transformation, otherwise the measurement would be biased. This symmetry
is equivalent to parity in the $z$-direction, effectuated by operators
$\hat\Pi_{z,{\rm S}} = \hat\sigma_x$ for the two-state system
and $\hat\Pi_{z,{\rm M}} = \exp\left({\rm i}\pi\sum_n \hat a_n^\dagger \hat a_n\right)$
for the meter \cite{Bru07}, so that the total system must be invariant under the transformation

\begin{equation} \label{spinmetsym}
\hat\Pi_z = \hat\Pi_{z,{\rm S}}\hat\Pi_{z,{\rm M}} =
\hat\sigma_x \exp\left({\rm i}\pi\sum_n \hat a_n^\dagger \hat a_n\right).
\end{equation}

\noindent
Indeed, it is readily verified that $\hat\Pi_{z,{\rm S}}^\dagger \hat H_{\rm S} \hat\Pi_{z,{\rm S}} =
\hat H_{\rm S}$, $\hat\Pi_{z,{\rm M}}^\dagger \hat H_{\rm M} \hat\Pi_{z,{\rm M}} = \hat H_{\rm M}$, and

\begin{align} \label{reflecsymsm}
\hat\Pi_z^\dagger \hat H_{\rm SM} \hat\Pi_z &=
\hat\Pi_{z,{\rm S}}^\dagger \hat\sigma_z \hat\Pi_{z,{\rm S}} \sum_n  g_n
\hat\Pi_{z,{\rm M}}^\dagger (\hat a_n^\dagger +
\hat a_n) \hat\Pi_{z,{\rm M}} \, \Theta(t) \nonumber\\
&= (- \hat\sigma_z) \left(- \sum_n  g_n
(\hat a_n^\dagger + \hat a_n) \right) \Theta(t) \nonumber\\
&= \hat H_{\rm SM}.
\end{align}

\noindent
Given this invariance, the Hilbert space of the total system decomposes into
two eigensubspaces of $\hat\Pi_z$,

\begin{equation} \label{paritydecomp}
\mathcal{H} = \mathcal{H}_+ \otimes \mathcal{H}_-,
\end{equation}

\noindent
$\mathcal{H}_+$ comprising symmetric, $\mathcal{H}_-$ antisymmetric states under
$\hat\Pi_z$. As the object (two-state) as well as the meter (boson) sector
of the total system can each be decomposed individually into an even and an odd subspace,
the parity subspaces decompose further into

\begin{equation} \label{sbparitydecomp}
\begin{split}
\mathcal{H}_+ &= \mathcal{H}_{{\rm S},+} \otimes \mathcal{H}_{{\rm M},+} \oplus
\mathcal{H}_{{\rm S},-} \otimes \mathcal{H}_{{\rm M},-}, \\
\mathcal{H}_- &= \mathcal{H}_{{\rm S},+} \otimes \mathcal{H}_{{\rm M},-} \oplus
\mathcal{H}_{{\rm S},-} \otimes \mathcal{H}_{{\rm M},+}.
\end{split}
\end{equation}

At the same time, both possible measurement outcomes, $|\!\!\downarrow\rangle$ as
well as $|\!\!\uparrow\rangle$, manifestly break the invariance under $z \to -z$ individually,
even if on average, the balance is equilibrated. In the framework of a unitary time evolution,
where the Hamiltonian as well as the initial state of the object are symmetric, the only
possible explanation left is that the asymmetry is introduced by the initial state of the meter.

Reconstructing the measurement in a unitary account of the full object-meter system
allows us to pursue the time evolution of the total state vector in continuous time.
Yet it is desirable, in order to compare with the standard view of quantum measurement,
 to record diagnostics that enable assessing the progress towards a definite
classical outcome. Two aspects are of particular significance for this purpose:
The approach of the spin component towards a pure state can be quantified in terms
of the von-Neumann entropy \cite{Neu18} of the reduced density operator

\begin{equation} \label{infospin}
I_{\rm S}(t) = -c\Tr_{\rm S}\bigl[\hat\rho_{\rm S}(t) \ln\bigl(\hat\rho_{\rm S}(t)\bigr)\bigr], \quad
\hat\rho_{\rm S}(t) = \Tr_{\rm M}\bigl[\hat\rho(t)\bigr],
\end{equation}

\noindent
or more specifically as its purity, $P_{\rm S}(t) = \Tr_{\rm S}\bigl[\hat\rho_{\rm S}^2(t)\bigr]$. Representing $\hat\rho_{\rm S}(t)$ as a Bloch vector ${\bf a} = (a_x,a_y,a_z)$,
$a_x := \frac{1}{2}\Tr(\hat\rho_{\rm S}\hat\sigma_x$ etc., the purity is reflected in its length,
$P_{\rm S}(t) = \frac{1}{2}(1 + |{\bf a}|^2)$. The asymmetry of the spin state
with respect to $z$-parity can be expressed as its polarization,

\begin{equation} \label{polaspin}
a_z(t) = \frac{1}{2} \bigl(\rho_{\uparrow\uparrow}(t) - \rho_{\downarrow\downarrow}(t)\bigr) =
\frac{1}{2} \bigl(\langle \uparrow\!\! | \hat\rho(t) | \!\!\uparrow \rangle -
\langle \downarrow\!\! | \hat\rho(t) | \!\!\downarrow\rangle \bigr), 
\end{equation}

\noindent
that is, as the vertical ($z$-) component of the Bloch vector.

\subsection{Simulating decoherence by finite heat baths}\label{sec43}

An essential condition to achieve an irreversible loss of coherence in a system
coupled to a macroscopic environment is that the spectrum of the environment,
be it composed of harmonic oscillators, spins \cite{CPZ05}, or other suitable microscopic
models, be continuous on the energy scales of the central system, or equivalently, that
the number $N$ of modes the environment comprises be large, $N \gg 1$.
As a general rule, based on energy-time uncertainty, recurrences
occur on a time scale $1/\Delta\omega$ if the spectrum exhibits structures
on the scale $\Delta\omega$. However, in the present context of a unitary
model for quantum measurement, it is more appropriate to avoid the limit $N \to \infty$.
Evidently, this can be achieved only if at the same time, irreversibility as a
hallmark of decoherence is sacrificed.

\begin{figure}[H]
\centering
\includegraphics[width=14 cm]{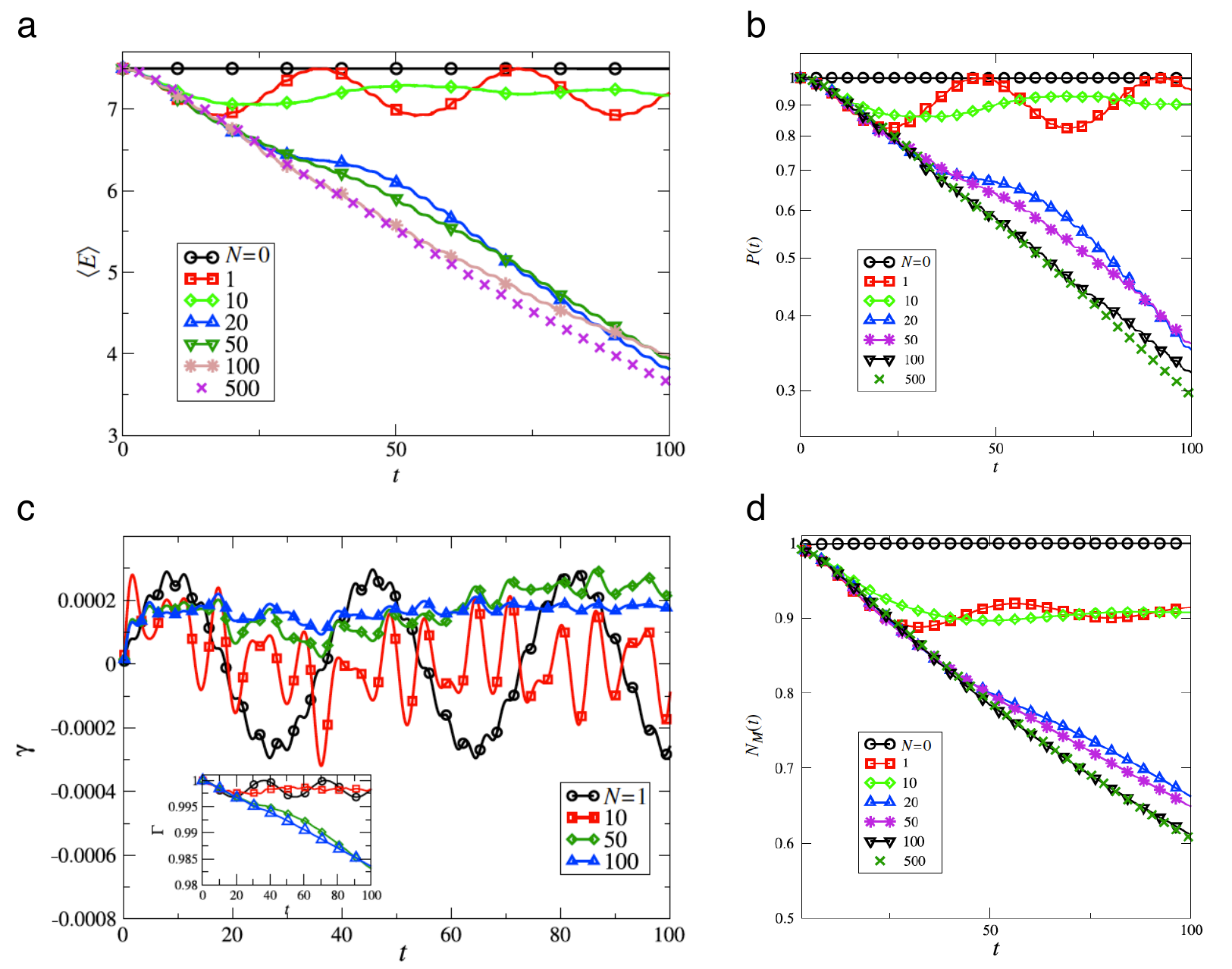}
\caption{Decoherence-like behaviour can be simulated by coupling a harmonic
oscillator to a reservoir comprising only a finite number $N$ of boson modes (harmonic
oscillators as well). The figure shows the time evolution of four diagnostics of decoherence
for different values of $N$, ranging from $N = 0$ (isolated central system) through
10, 20, 50, 100, through 500 (see legend). (\textbf{a}) Total energy in the central
system, showing a crossover from exponential to power-law decay for $N \geq 10$.
(\textbf{b}) Purity $P(t) = \Tr[(\rho_{\rm S}(t))^2]$. (\textbf{c}) Instantaneous dissipation
rate, i.e., ratio of effective friction force to time-dependent velocity (inset: total energy as
in panel (a)), for $N = 1$, 10, 50, 100.  (\textbf{d}) Degree of memory, measured as the
non-Markovianity $N_{\rm M}(t) = \frac{1}{t} \int_0^t {\rm d}t'\, |P(t')|$, $P(t)$ denoting 
the purity as depicted in panel (b). Reproduced from \cite{GBS14} with kind permission.}
\label{figfinbath}
\end{figure}

This price appears acceptable, though, as long as a faithful description of
the processes of interest is required only over a correspondingly large,
but finite time scale, as is the case, for example, in computational molecular
physics and in quantum optics. Numerical experiments simulating decoherence
with heat baths of finite Hilbert space dimension \cite{GKG10,Has11,GBS14} provide
convincing evidence that even with a surprisingly low number of bath modes, $N$ of
the order of 10, most relevant features of decoherence can be satisfactorily reproduced,
see Fig.\ \protect\ref{figfinbath}. This suggests to restrict the dimension of the meter sector
of the Hilbert space underlying the Hamiltonian (\ref{spinmetham}) accordingly
to a finite number $N$,

\begin{equation} \label{finmetham}
H = \frac{1}{2} \hbar\omega_0\hat\sigma_x +
\sum_{n=1}^N  g_n \hat\sigma_z (\hat a_n^\dagger + \hat a_n)\, \Theta(t) +
\sum_{n=1}^N \hbar\omega_n \left(\hat a_n^\dagger \hat a_n + \frac{1}{2}\right).
\end{equation}

\noindent
Like this, the Hamiltonian can be considered as a model of, e.g., a two-level atom
in a high-$Q$ microwave cavity \cite{RBH97}. The mode number $N$ thus
assumes the r\^ole of a central parameter of the model.

Concluding from the experience with similar models comprising finite baths
\cite{GKG10,GBS14}, the following scenarios appear plausible:
\begin{itemize}

\item For small values $N \gtrsim 1$, the time evolution comprises only a few,
but typically incommensurate, frequencies and should appear quasi-periodic.

\item Already for moderate numbers, say $N = O(10)$, the unitary model
will exhibit some characteristic features of quantum measurement. In particular,
a plausible scenario is that the object state approaches one of the pointer states
and remains in its vicinity for a longer time, before it may jump to another
(in the case of spin measurement, the opposite) pointer state. A similar
behaviour has indeed been observed for standard models of quantum optics
and solid-state physics and is known as ``collapses and revivals'' \cite{RBH97}.

\item For $N \gg 1$, the excursions of the object state away from pointer states
will become smaller and the frequency of switching episodes---spin flips in the case
of spin measurements---should reduce, that is, the times the object spends
close to a pointer state should grow very large. In particular, as soon as the object state
is sufficiently close to one of the pointer states, a behaviour reminiscent of the quantum
Zeno effect should emerge \cite{Zur82}.

\item With these intermediate stages, the limit $N \to \infty$ where a definite measurement
result is irreversibly achieved, while out of reach of finite-bath models proper, could still
be approached through of a continuous transition.

\end{itemize}

\noindent
In fact, a similar scenario has been envisaged for a model in the spirit of
quantum optics, representing the object by a two-state atom and meter and environment,
respectively, by two microwave cavities coupled through a waveguide \cite{RBH97}.

Of special interest is the opposite extreme, $N = 1$, as it allows us to study
some issues analytically that are no longer so directly accessible for higher values of $N$.
The Hamiltonian

\begin{equation} \label{spinbosonham}
H_{\rm sb} = \frac{1}{2} \hbar\omega_0\hat\sigma_x +
g \hat\sigma_z (\hat a^\dagger + \hat a)\, \Theta(t) +
\hbar\omega_1 \left(\hat a^\dagger \hat a + \frac{1}{2}\right).
\end{equation}

\noindent
also referred to as \emph{spin-boson Hamiltonian} or \emph{quantum Rabi model}
\cite{FG94,IGMS05}, is frequently employed as the standard model for two-level atoms
interacting with a bosonic field. It is often considered in a slightly simplified version: If a
rotating-wave approximation is applied that excludes double excitation or de-excitation
processes (generated by $\hat\sigma_+ \hat a^\dagger$ or $\hat\sigma_- \hat a$), the interaction
term reduces to $\hat H_{\rm SM} = g(\hat\sigma_+ \hat a + \hat\sigma_- \hat a^\dagger)$,
denoting $\hat\sigma_\pm := \frac{1}{2}(\hat\sigma_x \mp {\rm i}\hat\sigma_y)$. With this
modification, the spin-boson Hamiltonian is also known as \emph{Jaynes-Cummings model}.
The emblematic feature exhibited by spin-boson systems are \emph{Rabi oscillations},
oscillations of the two-state system between its lower and its upper level with a frequency
proportional to the coupling $g$. A further simplification of Eq.\ (\ref{spinbosonham}),
often called \emph{semi-classical Rabi model}, replaces the coupling to the boson mode
with frequency $\omega_1$ by an external driving with the same frequency \cite{GH92,BCB&16},
$H_{\rm scl} = \frac{1}{2} \hbar\omega_0\hat\sigma_x + g \hat\sigma_z \cos(\omega_1 t)$.

With the Hamiltonian (\ref{spinbosonham}), it is straightforward to specify the
parity eigensubspaces referred to in Eq.\ (\ref{sbparitydecomp}). The even
eigenspace comprises states of the form

\begin{equation} \label{sbevenpar}
\begin{split}
\bigl\lvert \Psi_{++}\bigr\rangle &=
\frac{1}{sqrt{2}} \bigl(|\!\!\downarrow\rangle + |\!\!\uparrow\rangle\bigr)
\sum_{\alpha=0}^\infty c_{2\alpha} |2\alpha\rangle \quad{\rm or} \\
\bigl\lvert \Psi_{--}\bigr\rangle &=
\frac{1}{\sqrt{2}} \bigl(|\!\!\downarrow\rangle - |\!\!\uparrow\rangle\bigr)
\sum_{\alpha=0}^\infty c_{2\alpha+1} |2\alpha+1\rangle,
\end{split}
\end{equation}

\noindent
the odd subspace is spanned by states of the form

\begin{equation} \label{sboddpar}
\begin{split}
\bigl\lvert \Psi_{+-}\bigr\rangle &= \frac{1}{\sqrt{2}} \bigl(|\!\!\downarrow\rangle +
| \!\!\uparrow\rangle\bigr)
\sum_{\alpha=0}^\infty c_{2\alpha+1} |2\alpha+1\rangle \quad {\rm or} \\
\bigl\lvert \Psi_{-+}\bigr\rangle &=
\frac{1}{\sqrt{2}} \bigl(|\!\!\downarrow\rangle - |\!\!\uparrow\rangle\bigr)
\sum_{\alpha=0}^\infty c_{2\alpha} |2\alpha\rangle.
\end{split}
\end{equation}

Numerical results for the quantum dynamics, generated both by the Jaynes-Cummings
model \cite{Gea91} and by the complete spin-boson Hamiltonian \cite{FG94,IGMS05}, in a
parameter regime relevant for the present modelling, in particular for strong coupling,
exist already and are consistent with the expectations pointed out here.
For the present application to quantum measurement, there is no obvious justification
for a rotating-wave approximation. With the full Hamiltonian (\ref{spinbosonham}),
the von-Neumann equation for the density operator, ${\rm i}\hbar\, {\rm d}\hat\rho/{\rm d}t =
[H_{\rm sb},\hat\rho]$ is readily evaluated at $t = 0$ (App.\ \ref{inievsbham}).
Assuming an initial state as in Eq.\ (\ref{spinmetini}), factorizing into a Schr\"odinger cat
for the two-state system and an arbitrary superposition of boson excitations,

\begin{equation} \label{sbinistate}
|\Psi_\pm(0)\rangle = \frac{1}{\sqrt{2}} \bigl(|\!\!\downarrow\rangle \pm |\!\!\uparrow\rangle\bigr)
\sum_{\alpha=0}^\infty c_\alpha |\alpha\rangle
\end{equation}

\noindent
the evolution equation for the reduced two-state density operator at $t = 0$ reads

\begin{equation} \label{sbinirho1sttdev}
\frac{{\rm d}}{{\rm d}t} \hat\rho_{\rm S}(t)\Bigr\rvert_{t=0} =
\pm g \hat\sigma_y \sum_{\alpha=0}^\infty \sqrt{\alpha+1}\, {\rm Re}(c_{\alpha+1}c_\alpha^*) .
\end{equation}

\noindent
For the initial polarization $a_z = \frac{1}{2}\langle \hat\sigma_z \rangle$, this means

\begin{equation} \label{sbinipol1sttdev}
\frac{{\rm d}}{{\rm d}t} a_z(t)\Bigr\rvert_{t=0} =
\frac{1}{2} \bigl(\dot{\hat\rho}_{\uparrow\uparrow}(0) -
\dot{\hat\rho}_{\downarrow\downarrow}(0)\bigr) = 0.
\end{equation}

\noindent
That is, to leading order, the state vector starts rotating around the $z$-axis of the Bloch sphere,
but does not leave the equator. However, going to the second time derivative,

\begin{align} \label{sbinirho2ndttdev}
\frac{{\rm d}^2}{{\rm d}t^2}\, \hat\rho_{\rm S}(t)\Bigr\rvert_{t=0} =
&\pm 2g \sum_{\alpha=0}^\infty \sqrt{\alpha+1}\,\left(
\omega_0\hat\sigma_z {\rm Re}(c_{\alpha+1}c_\alpha^*) +
\omega_1\hat\sigma_y {\rm Im}(c_{\alpha+1}c_\alpha^*) \right) \nonumber\\
&\mp 2g^2\hat\sigma_x \sum_{\alpha=0}^\infty \left(|c_\alpha|^2(2\alpha+1) +
\sqrt{(\alpha+1)(\alpha+1)}\,{\rm Re}(c_{\alpha+2}c_\alpha^*)\right),
\end{align}

\noindent
one finds

\begin{equation} \label{sbinipol2ndttdev}
\frac{{\rm d}^2}{{\rm d}t^2}\, a_z(t)\Bigr\rvert_{t=0} =
\frac{1}{2} \bigl(\ddot{\hat\rho}_{\uparrow\uparrow}(0) -
\ddot{\hat\rho}_{\downarrow\downarrow}(0)\bigr) =
\pm 2g\omega_0 \sum_{\alpha=0}^\infty \sqrt{\alpha + 1}\,{\rm Re}(c_{\alpha+1}c_\alpha^*).
\end{equation}

\begin{figure}[H]
\centering
\includegraphics[width=7 cm,angle=270]{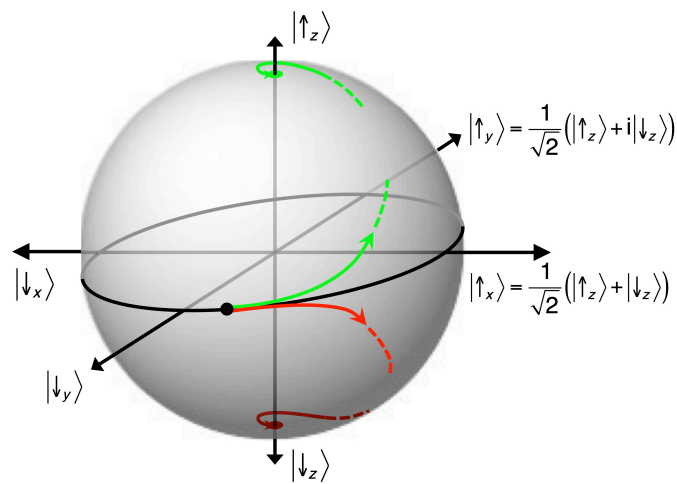}
\caption{Spin measurement on the Bloch sphere. The quantum dynamics of
spin measurements is dominated by two ``pointer states'', eigenstates of
the measured operator $\hat\sigma_z$, i.e., $|\!\!\uparrow_z\rangle$ and
$|\!\!\downarrow_z\rangle$, represented on the Bloch sphere as North
(green dot) and South pole (red dot). Owing to the quantum Zeno effect, they
attract nearby states of the measured system. At the same time, the short-time
evolution of the measured spin for a meter comprising only a single boson mode,
Eq.\ (\protect\ref{sbinipol2ndttdev}), suggests that a state initiated on the equator
of the Bloch sphere (black dot), besides rotating around the equator, will tend
towards one of the poles, depending on the initial state of the meter boson mode.}
\label{figbloch}
\end{figure}   

This result indicates that to second order in time, a state prepared as a Schr\"odinger cat
with respect to vertical spin will exhibit polarization if the initial state of the boson
fulfills a specific condition. The terms in the sum over $\alpha$ in Eq.\ (\ref{sbinipol2ndttdev})
only contribute if not all products $c_{\alpha+1}c_\alpha^*$ of two subsequent expansion
coefficients vanish. It has an obvious interpretation in terms of symmetry: The boson components
in the eigensubspaces of the parity operator $\hat\Pi_z$, Eqs.\ (\ref{sbevenpar},\ref{sboddpar}),
are characterized by encompassing exclusively even or exclusively odd components
of each sector, spin and boson, of the total system. The condition
$c_{\alpha+1}c_\alpha^* \neq 0$ for the boson sector therefore implies that the initial
state \emph{of the meter} must not belong to either one of the two eigensubspaces, hence
must break $z \to -z$ parity, while the initial state of the spin itself has to remain unbiased.

If this preliminary finding is combined with the quantum Zeno effect (Sect.\ \ref{sec41}),
a scenario emerges where initial states, unbiased as to spin polarization, will
tend to move away from the equator of the Bloch sphere, the attraction basin boundary
between spin-up and spin-down, in a direction depending on the initial state of the meter,
to become increasingly attracted by that pole of the Bloch sphere they are already
approaching, see Fig.\ \protect\ref{figbloch}.

\subsection{A classical analogue of spin measurement}\label{sec44}

Following a similar research program as in quantum chaos, comparing quantum dynamics
to its closest classical analogue, it would be tempting to study the unitary model for spin
measurement sketched above in some appropriate classical limit. For the boson sector,
no approximations are even necessary, as the heat bath composed of harmonic oscillators
is its own classical limit. The two-state system representing the measurement object,
however, is located in the opposite, the extreme quantum regime. A classical limit
in the formal sense does not exist or it. However, a model, closely analogous in many
respects to a spin measurement, can be conceived that already provides relevant insights.

\begin{figure}[H]
\centering
\includegraphics[width=13 cm]{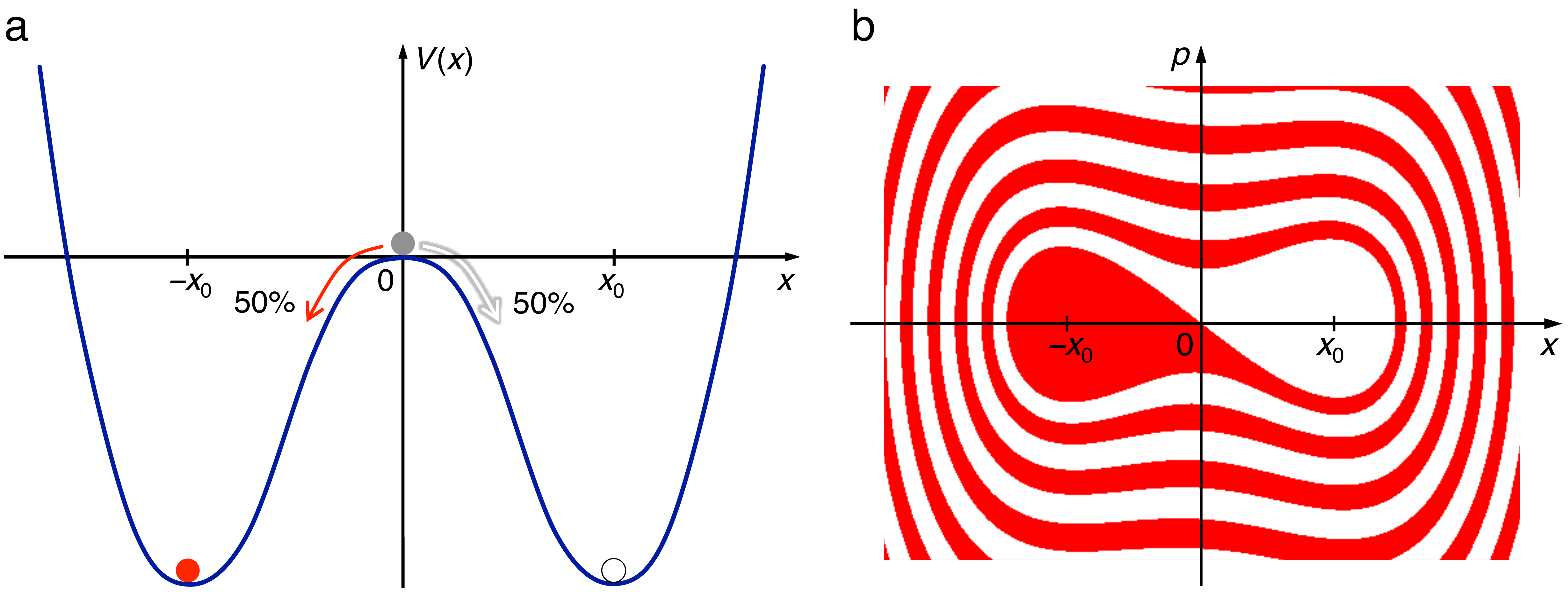}
\caption{Damped motion in a symmetric double-well potential. In a quartic potential
with a symmetric parabolic barrier (\textbf{a}), Eq.\ (\protect\ref{dwham}), from
an initial state with zero momentum in the unstable equilibrium position on top of the barrier, 
random impacts from an environment will send the system with equal probability
towards one of the potential minima at $\pm x_0 = \pm \sqrt{a/b}$. In the presence of friction,
Eq.\ (\protect\ref{dwdissnewton}), it will come to rest, once transient oscillations are damped out,
in that well which it approached initially, giving rise to (\textbf{b}) basins of attraction associated
to either one of the wells. The parameter values of Eq.\ (\protect\ref{dwdissnewton}) are
$a = 0.25$, $b = 0.01$, $\gamma = 0.04$.}
\label{figdoublewell}
\end{figure}   

The fact that, in the limit of a quasi-continuous heat bath, the two opposite
pointer states act as attractors in Hilbert space suggests to compare them
with a bistable classical system. A paradigm for bistability is a double well potential,
say a symmetric quartic double-well with a parabolic barrier
(Fig.\ \protect\ref{figdoublewell}a), given by the Hamiltonian

\begin{equation} \label{dwham}
H_{\rm S}(p_{\rm S},x_{\rm S}) = \frac{p_{\rm S}^2}{2m_{\rm S}} +V_{\rm S}(x_{\rm S}), \quad
V_{\rm S}(x_{\rm S}) = - \frac{a}{2} x_{\rm S}^2 + \frac{b}{4} x_{\rm S}^4
\end{equation}

\noindent
If the heat bath takes the same form as in Eq.\ (\ref{finmetham}),

\begin{equation} \label{clhbham}
H_{\rm M}({\bf p},{\bf x}) = \sum_{n=1}^N \left(\frac{p_n^2}{2m_n} +
m_n\omega_n^2 \frac{x_n^2}{2} \right),
\end{equation}

\noindent
and the interaction is modelled, as in the quantum case, as a linear position-position
coupling,

\begin{equation} \label{dwhbham}
H_{\rm SM}(x_{\rm S},{\bf x}) = g x_{\rm S} \sum_{n=1}^N x_n,
\end{equation}

\noindent
the total Hamiltonian takes the form

\begin{equation} \label{cldwham}
H(p_{\rm S},x_{\rm S},{\bf p},{\bf x}) =
H_{\rm S}(p_{\rm S},x_{\rm S}) + H_{\rm SM}(p_{\rm S},x_{\rm S},{\bf p},{\bf x}) +
H_{\rm M}({\bf p},{\bf x}).
\end{equation}

\noindent
It is evidently symmetric under the parity operation $P_{x_{\rm S},{\bf x}}$:
$(p_{\rm S},x_{\rm S}) \to (-p_{\rm S},-x_{\rm S})$,
$({\bf p},{\bf x}) \to (-{\bf p},-{\bf x})$.
An initial condition that comes as close as possible to Eq.\ (\ref{spinmetini}),
in particular to the Schr\"odinger cat state for the spin, would combine the
double-well system prepared at rest in the unstable equilibrium position on top of
the barrier (call it a ``Buridan's ass state''), $p_{\rm S}(0) = 0$, $x_{\rm S}(0) = 0$,
with an arbitrary initial condition of the heat bath oscillators, ${\bf p}(0) = {\bf p}_0$,
${\bf x}(0) = {\bf x}_0$. Suitable metaphors for this model are an inverted pendulum
or a pencil, initially balanced exactly vertically, tip down on a surface, exposed
to the impinging molecules of the surrounding medium. 

As in the quantum model, the number of degrees of freedom $N$ of the heat bath
is a decisive parameter. Already for $N = 1$, chaotic behaviour is expected
for the coupled system. In the limit $N \to \infty$, it should approach an
irreversible dynamics, characterized by dissipation. With a position-position coupling
as in Eq.\ (\ref{dwhbham}) and under similar conditions for the spectrum of the heat bath,
it will take the form of Ohmic friction (proportional to the velocity of the damped
degree of freedom). For the object, this would imply an equation of motion such as

\begin{equation} \label{dwdissnewton}
m_{\rm S} \ddot{x}_{\rm S} = -\lambda\dot{x}_{\rm S} + ax_{\rm S} - bx_{\rm S}^3,
\end{equation}

\noindent
with a friction coefficient $\lambda$ that depends on the microscopic coupling $g$
and the spectrum of the heat bath. For moderate values of $\gamma$, the system
will fall from the top of the barrier into one of the wells and, after oscillations within
that well have faded out, remain at rest in that well. As the Hamiltonian as well as
the initial state of the object are parity symmetric, it is the initial conditions of
the heat bath oscillators which determine into which one of the two wells the
object will fall. While the boundary between the basins of attraction of the two
wells (Fig.\ \protect\ref{figdoublewell}b) passes exactly through the origin of phase space,
that is, through the supposed initial state $p_{\rm S}(0) = 0$, $x_{\rm S}(0) = 0$, it
becomes fuzzy in the presence of the environment and is displaced slightly towards
one of the two wells, depending on the initial condition of the environment.
For the inverted pendulum alluded to above this means that it will tilt over
in a direction determined by the impact of the molecules of the surrounding medium.

A similar model for a classical binary ``random'' (but in fact deterministic) process,
a coin toss, has been analyzed in all detail in Ref.\ \cite{DHM07}. Diaconis \emph{et al.}
construct a phase-space plot of the basin boundaries separating initial conditions of
the coin that lead to either one of the two alternative final outcomes ``head'' and ``tail'',
and which shows a marked structure of alternating fine fringes corresponding to these 
final conditions. While in the case of coin tosses, the sensitive dependence on the initial
condition of the coin itself serves as random generator, it is microscopic details of
the initial state of the environment that contribute the required entropy in the above
classical model of spin measurement.

\subsection{Perspectives}\label{sec45}

A unitary account of quantum measurements with random outcome, as outlined
in this section, is presently being worked out. Starting from the analytical framework presented here,
it requires massive numerical calculations. The quantum model with finite mode number $N$
can be evaluated in numerical simulations following a similar strategy as in the cited work
on finite heat baths in optics and quantum molecular dynamics. The classical model of
a bistable measurement process gives rise to sets of coupled Hamiltonian equations
of motion that can be integrated using symplectic solvers.

In both cases, the immediate objective is to increase the mode number as far as possible,
in order to come close to an irreversible behaviour, at least on time scales larger than
all characteristic times of the object. The scenario sketched above for sufficiently
high values of $N$ is a plausible expectation, based on arguments involving analogies
and extrapolating known results. If it proves correct, a strong point would be made in
favour of an alternative view of quantum randomness. Instead of interpreting it as a
fundamental element of quantum mechanics that remains out of reach of its standard
formalism, it would integrate it in a similar category of in principle deterministic,
but practically incalculable many-body phenomena as, e.g., classical thermal fluctuations or
Brownian motion. The randomness manifest in spin measurement would be revealed as
analogous to quantum vacuum state fluctuations, amplified to macroscopic observability
and frozen in a lasting record.

If, on the contrary, quasi-stable measurement outcomes should not be achieved
even for values $N \gg 1$, this would provide strong evidence that quantum
measurement involves a class of randomness that is fundamentally different
from all other known sources of stochastic behaviour and will remain an alien
element in the framework of quantum mechanics. It would constitute a
compelling reason to explore unprecedented ideas---after all, the general problems
pointed out above, concerning the information balance and spontaneous
symmetry breaking in the ``second collapse'', remain pressing.

An unexpected but important consequence of this view is that it effectively merges
the ``first'' and the ``second'' collapse of the wavepacket into a single unitary
process. In this way, it avoids the conceptually inconvenient detour from a pure
initial state (a Schr\"odinger cat) to a mixture, after the first collapse, and back to
a pure state (a definite measurement result) and in particular complies with
entropy conservation throughout the entire measurement.

Besides this central message, a unitary account of quantum measurement
has various additional testable implications, which substantiate its strength
and facilitate its falsification:

\begin{itemize}

\item The approach of the object state to one of the pointer states, as final result
of the measurement, will never be complete. In the limit $N \to \infty$, the discrepancy
is expected to become arbitrarily small, but the postulate of pure states resulting
from quantum measurement cannot be accomplished literally.

\item Owing to the unavoidable entanglement between object and meter,
the initial state of the meter does not only affect the final state of the object
(the result), the state of the object upon leaving the apparatus also leaves
a trace in the meter, which can then be probed by the following measurement. This
implies the possibility of correlations between subsequent spin measurements,
otherwise incompatible with their randomness, if their separation in time is extremely short.

\item Spin measurements on systems prepared as Schr\"odinger cats with
respect to the measured spin component, a paradigm of quantum randomness,
are in the focus of this work. Notwithstanding, also ``redundant'' measurements,
performed on systems that are prepared already with a definite polarization in the
measured direction, are of interest in this context: The existence of a back-action of
the meter on the object implies that even in the case of redundant measurements,
albeit with very low probability, the measurement process could alter
the spin polarization---trigger a spin flip---so that the result would not coincide
with the state of the spin upon entering the apparatus.

\item The approach outlined herein emphasizes the relevance of
the meter state for the measurement outcome. Besides its initial state proper,
this includes also characteristic properties of the meter, such as its eigenenergy
spectrum and the way it couples to the object. Unconventional features, achieved
by some special design, may then be reflected in unexpected features also
of the statistics of the measurement results. If, for example, the ``meter''
is represented by a microwave cavity, as is often the case in quantum optics,
particular structures in the cavity spectrum could have an observable effect
on the measurement results.

\item In state-of-the-art laboratory experiments on quantum randomness
\cite{BK&18}, photons in counter-rotating polarization states replace the
spins traditionally used as qbits in this context. It appears possible and tempting
to work out the theory developed here so as to apply it to photon experiments.

\end{itemize}

Random spin measurements are almost invariably discussed in a special
context where indeed they play a crucial r\^ole: Einstein-Podolsky-Rosen (EPR)
experiments \cite{EPR35,Bel64,ADR82}. This issue has deliberately been avoided
here, as it is charged with misleading connotations. In particular,  in EPR experiments, 
quantum randomness is not only inextricably connected to nonlocality, it is even discussed
as depending on it as on a necessary condition \cite{BK&18}. The present approach,
however,  is unrelated to this question, and it is not intended to contribute in any sense to
the long-standing debate around nonlocality and hidden-variable approaches.
The attempt to understand quantum randomness within the established framework of
non-relativistic unitary quantum time evolution is intended to fill one of the last gaps where 
it could appear incomplete, but not to question it, let alone replace it with hypotheses reviving
classical locality. 

Yet it cannot be denied that it has implications also for the
interpretation of EPR experiments. Should it be the case that the meter
has an impact on individual spin measurements, how then can spontaneous
correlations arise between simultaneous measurements on spin pairs with
a space-like separation? A possible answer could lie in the assumption that
the nonlocal common state of the measured spin pair is coupled to, thus gets
entangled with, the initial states of \emph{both} meters, in a similar way as a single spin
supposedly probes the state of a single meter. The coupling to the nonlocal
spin state would then indirectly also entangle the states of the two apparatuses with
one another, even if they remain separated by a space-like distance. However,
this question should be relegated to future research as a particularly intriguing subject,
to be addressed once the basic questions raised in this proposal have been settled.

\section{Conclusions}\label{sec5}

The present report spans a wide arc, from minimalist models of chaos inspired by card
shuffling, through pseudo-chaotic behaviour in pixelated spaces, through the quantum death
of classical chaos, through spin measurement. These diverse subjects do have a common
denominator. They allow to peek, from a macroscopic observation platform, into
details of information processing on the smallest scales, directing attention to a few
essential aspects: fundamental limits of total information supply and storage density
on these scales, ``vertical'' information currents interchanging entropy with large scales,
``horizontal" exchange of entropy with adjacent degrees of freedom of the environment.

They are relevant in particular for an understanding of stochastic processes,
collectively perceived as ``randomness'', on the macroscopic level. The analysis
presented here supports the view that they form windows to the microscopic world,
exceptional points where information is not dumped into, but lifted up from
small scales, like hot magma ascending from the Earth's mantle that reaches the surface
in volcanic eruptions. While this idea may be little more than a helpful metaphor
in the context of classical chaos, it suggests surprising consequences
if applied to a seemingly unrelated field, quantum measurement. The randomness
generated in quantum measurement can be seen in a similar spirit as  
resulting from an instability of the coupled object-meter system as it evolves
towards alternative measurement results. If it can be evidenced that also here,
it is information exchanged with the meter that becomes manifest in the
measurement result, that would suggest an interpretation of quantum
randomness as amplified vacuum fluctuations, rather than an irreducible
fundamental feature of quantum systems.

An interpretation and extrapolation of quantum chaos in this sense is but a single example
of the fruitfulness of studying quantum phenomena in terms of information currents.
This approach, originating in and inspired by the success of quantum information
science applied to computing, is developing into an active research area of its own right,
with applications in quantum optics, quantum many-body physics, and other areas
waiting to be explored.

While entropy and information currents have proven invaluable tools to understand 
classical and quantum chaos, the discussion of randomness in quantum measurement
reveals a significant shortcoming of quantum entropy as an analytical instrument:
It is insensitive to the difference between ordered strings and random strings.
Intuitively,  a structural criterion for randomness should also be reflected in a suitable
entropy measure for quantum processes, as it is indeed addressed on the classical
level, notably in the context of algorithmic complexity \cite{Sol64,Kol65,Cha66,Zur89}.

\funding{A research grant from \emph{Fundaci\'on para la Promoci\'on de la Investigaci\'on y la Tecnolog\'\i a} (FPIT) of \emph{Banco de la Rep\'ublica de Colombia} (grant number 4050)
is gratefully acknowledged.}

\acknowledgments{The author enjoyed the hospitality of the Department of Physics of Complex
Systems at the Weizmann Institute of Science (Rehovot, Israel) and at the Institute for
Theoretical Physics at Technical University of Dresden (Dresden, Germany) during various
research stays, where part of the work reported here has been performed, and gratefully
acknowledges stimulating discussions with his hosts at these institutions, Uzy Smilansky
and Walter Strunz, resp., as well as with Frank Gro\ss mann (TU Dresden), Roee Ozeri
(Weizmann Institute), and Carlos Viviescas (Universidad Nacional de Colombia).
I am indebted to Santiago Pe\~na for preparing the plot of attraction basins of a bistable
system that underlies Fig.\ \protect\ref{figdoublewell}b.}

\appendixtitles{yes} 
\appendixsections{multiple} 
\appendix

\section{Entropy conservation under classical canonical transformations} \label{clentrocons}

For a classical mechanical system comprising $f$ degrees of freedom, specify the state as a probability
density function

\begin{equation} \label{rhofbyfdef}
\rho:\;  \mathbb{R}^{2f} \to \mathbb{R}^+,\;
\mathbb{R}^{2f}\ni {\bf r} \mapsto \rho({\bf r}) \in \mathbb{R}^+, \quad
\int {\rm d}^{2f}r\, \rho({\bf r}) = 1.
\end{equation}

\noindent
In the absence of birth and death processes, ${\rm d}\rho({\bf r},t)/{\rm d}t = 0$,
it evolves in time according to the Liouville equation \cite{Gol80}

\begin{equation} \label{rholiouville}
\frac{\partial}{\partial  t}\, \rho({\bf r},t) = \{H({\bf r},t),\rho({\bf r},t)\},
\end{equation}

\noindent
$\{H({\bf r},t),\rho({\bf r},t)\}$ denoting the Poisson bracket with the Hamiltonian $H({\bf r},t)$.
For the evolution over finite times, say from $\rho({\bf r}',t')$ to $\rho({\bf r}'',t'')$,
that means that the density is conserved along a trajectory or flow line, 

\begin{equation} \label{rhotrajectory}
\rho({\bf r}'',t'') = \rho\bigl({\bf r}'({\bf r}''),t'\bigr) = \rho\bigl(\hat{\bf F}^{-1}(t'',t'){\bf r}'',t'\bigr),
\end{equation}

\noindent
where the operator-valued vector function $\hat{\bf F}(t'',t')$ maps phase-space points
${\bf r}'$ at time $t'$ along their trajectory till $t''$. Conversely, $\hat{\bf F}^{-1}(t'',t')$
traces phase-space vectors back along their trajectory from $t''$ to $t'$.

For a state given by a continuous probability density at time $t$, the classical information
can be defined as

\begin{equation} \label{rhophasespaceinfo}
I(t) = -c \int {\rm d}^{2f}r\, \rho({\bf r},t) \ln\bigl(d_A^f \rho({\bf r},t)\bigr).
\end{equation}

\noindent
The constant $c$ fixes the units of information, $d_A$ is the resolution in units of action
in two-dimensional phase space, given for example by the accuracies $d_x$ of length
and $d_p$ of momentum measurements, $d_A = d_x d_p$. In order to relate the information
at time $t''$ to that at an earlier or later time $t'$, we can refer to the evolution of
the density over a finite time interval, Eq.\ (\ref{rhotrajectory}),

\begin{align}
I(t'') &= -c \int {\rm d}^{2f}r''\, \rho({\bf r}'',t'') \ln\bigl(d_A^f \rho({\bf r}'',t'')\bigr) \nonumber\\
&=  -c \int {\rm d}^{2f}r''\, \rho(\hat{\bf F}^{-1}(t'',t'){\bf r}'',t')
\ln\Bigl(d_A^f \rho\bigl(\hat{\bf F}^{-1}(t'',t'){\bf r}'',t'\bigr)\Bigr). \label{rhofinitiminfo}
\end{align}

\noindent
It suggests itself to change the integration variable from the ``new'' phase-space coordinate
${\bf r}''$ to the ``old'' one ${\bf r}'$, involving the Jacobian determinant
$\det(\partial {\bf r}''/\partial {\bf r}')$. The $(2f\times 2f)$-matrix $M$,
also known as stability matrix, linearizes the transformation $\hat{\bf F}$,

\begin{equation} \label{stabilitymatrix}
M = \frac{\partial {\bf r}''}{\partial {\bf r}'} =
\frac{\partial}{\partial {\bf r}'}\, \hat{\bf F}(t'',t'){\bf r}'.
\end{equation}

\noindent
In the framework of Hamiltonian mechanics, $\hat{\bf F}$ must be canonical,
which requires that $M$ complies with the symplectic condition $M^{\rm t}JM = J$,
$J$ denoting the symplectic unit matrix \cite{Gol80}. For the Jacobian, it means that
$\bigl(\det(M)\bigr)^2 = 1$. This allows to rewrite the integration in Eq.\ (\ref{rhofinitiminfo}) as,

\begin{align}
I(t'') &=  -c \int {\rm d}^{2f}r'\,
|\det(M)|\, \rho({\bf r}',t') \ln\bigl(d_A^f \rho({\bf r}',t')\bigr) \nonumber\\
&=  -c \int {\rm d}^{2f}r'\,  \rho({\bf r}',t') \ln\bigl(d_A^f \rho({\bf r}',t')\bigr) \nonumber\\
&= I(t'). \label{clinfocons}
\end{align}

\noindent
The conservation of information in classical Hamiltonian dynamics, manifest in
Eq.\ (\ref{clinfocons}), evidently is a lemma of symplectic phase-space volume conservation 
under canonical transformations \cite{Gol80}. It is also as general: For example, it extends
unconditionally also to systems driven by a time-dependent external potential force,
which typically do \emph{not} conserve energy.

\section{Entropy conservation under quantum unitary time evolution} \label{qmentrocons}

As the most general measure of the information content of the state of a quantum system,
described by the density operator $\hat\rho(t)$, define the von-Neumann entropy,
\begin{equation} \label{vnentropy}
I(t) = -c \Tr[\hat\rho(t) \ln(\hat\rho(t))].
\end{equation}
Based on the density operator, this definition readily covers time evolutions that include
incoherent processes, such as dissipation or measurement. In the special case of a
unitary time evolution, generated by a Hamiltonian $\hat H(t)$ (that may well depend
on time), the density operator evolves according to the von-Neumann equation \cite{CDL77}

\begin{equation} \label{vnequation}
\frac{{\rm d}}{{\rm d}t}\, \hat\rho(t) = \frac{-{\rm i}}{\hbar} \bigl[\hat H(t),\hat\rho(t)\bigr].
\end{equation}

\noindent
The evolution over a finite time, from $\hat\rho(t')$ to $\hat\rho(t'')$, generated by
Eq.\ (\ref{vnequation}),

\begin{equation} \label{densopunev}
\hat\rho(t'') = \hat U(t'',t')\, \hat\rho(t')\, \hat U^\dagger (t'',t'),
\end{equation}

\noindent
is mediated by the unitary time evolution operator

\begin{equation} \label{timevop}
\hat U(t'',t') = \hat{\mathcal{T}} \exp\left(\frac{-{\rm i}}{\hbar} \int_{t'}^{t''} {\rm d}t\, \hat H(t) \right),
\end{equation}

\noindent
where the operator $\hat{\mathcal{T}}$ effectuates time ordering.

Combining Eq.\ (\ref{vnentropy}) with (\ref{timevop}), the von-Neumann entropy \cite{NC00}
is found to evolve from $t'$ to $t''$ as

\begin{align}
I(t'') &=  -c \Tr[\hat\rho(t'') \ln(\hat\rho(t''))] \nonumber\\
&=  -c \Tr\left[\hat U(t'',t')\, \hat\rho(t')\, \hat U^\dagger (t'',t')
          \ln\Bigl(\hat U(t'',t')\, \hat\rho(t')\, \hat U^\dagger (t'',t')\Bigr)\right] \label{qminfoinit}.
\end{align}

\noindent
In order to evaluate the trace, expand the operator-valued log function
in a Taylor series around the identity $\hat I$,
$\ln(\hat I + \hat x) = \sum_{n=1}^\infty a_n \hat x^n$, $a_n =
\ln^{(n)}(1)/n! = (-1)^{n-1} / n$,

\begin{equation}
I(t'') =  -c \Tr\left[\hat U(t'',t')\, \hat\rho(t')\, \hat U^\dagger (t'',t')
               \sum_{n=1}^\infty a_n
               \bigl(\hat U(t'',t')\, \hat\rho(t')\, \hat U^\dagger (t'',t') - \hat I\bigr)^n \right].
\end{equation}

\noindent
Permuting factors under the trace and eliminating intermediate products
$\hat U^\dagger (t'',t') U(t'',t') = \hat I$,

\begin{align}
I(t'') &=  -c \Tr\left[\hat\rho(t')\, \sum_{n=1}^\infty a_n \, \hat U^\dagger (t'',t')
      \bigl(\hat U(t'',t')\, (\hat\rho(t') - \hat I) \hat U^\dagger (t'',t') \bigr)^n
      \hat U(t'',t') \right] \nonumber\\
&=  -c \Tr\left[\hat\rho(t')\, \sum_{n=1}^\infty
a_n (\hat\rho(t') - \hat I)^n \right] \label{qminfocont},
\end{align}

\noindent
the sum under the trace recomposes to
\begin{equation} \label{qminfocons}
I(t'') =  -c \Tr[\hat\rho(t') \ln(\hat\rho(t'))] = I(t').
\end{equation}

\noindent
The decisive argument in this derivation is evidently that unitary transformations
leave the trace of transformed operators invariant, in direct analogy to the conservation
of phase-space volume under canonical transformations that guarantees entropy
conservation in classical Hamiltonian dynamics, cf.\ App.\ \ref{clentrocons}.

\section{Initial time evolution for the spin-boson Hamiltonian with a single boson mode}
\label{inievsbham}

For the spin-boson Hamiltonian with a ``heat bath'' comprising only a single harmonic
oscillator, cf.\ Eq.\ (\ref{spinbosonham}),

\begin{equation} \label{spinbosonhamapp}
H_{\rm sb} = \frac{1}{2} \hbar\omega_0\hat\sigma_x + g \hat\sigma_z (\hat a^\dagger + \hat a)\,
\Theta(t) + \hbar\omega_1 \left(\hat a^\dagger \hat a + \frac{1}{2}\right),
\end{equation}

\noindent
a few key quantities, such as the reduced density operator of the spin sector and
its polarization, are analytically accessible at the initial time $t = 0$.

Prepare the boson mode in an arbitrary superposition of eigenstates,

\begin{equation} \label{sbiniboson}
| \psi_{\rm M}(0) \rangle = \sum_{\alpha = 0}^\infty c_\alpha |\alpha\rangle, \quad
\sum_{\alpha = 0}^\infty |c_\alpha|^2 = 1
\end{equation}

\noindent
and the spin in a Schr\"odinger cat state 

\begin{equation} \label{sbinispin}
\vert \psi_{\rm S,ini} \rangle =
\frac{1}{\sqrt{2}} \bigl(| \!\!\downarrow \rangle \pm |\!\!\uparrow \rangle\bigr).
\end{equation}

\noindent
This amounts to an initial condition of the reduced density operator

\begin{equation} \label{sbinispinrhoop}
\hat\rho_{\rm S}(0) = \Tr_{\rm M}\bigl(\hat\rho(0)\bigr) =
\frac{1}{2} \bigl(I_0 \pm \hat\sigma_x\bigr),
\end{equation}

\noindent
i.e., in the representation of the eigenstates of $\hat\sigma_z$,

\begin{equation} \label{sbinispinrhomat}
\rho_{\rm S}(0) = \frac{1}{2} \begin{pmatrix} 1 & \pm 1 \\ \pm 1 & 1 \end{pmatrix}.
\end{equation}

\noindent
Evidently, it represents a pure state, $\bigl(\hat\rho_{\rm S}(0)\bigr)^2 = \hat\rho_{\rm S}(0)$.

Its first time derivative is obtained immediately from the von-Neumann equation,

\begin{align} \label{sbinispinrhoopddt}
\frac{{\rm d}}{{\rm d}t}\, \hat\rho_{\rm S}(t)\Bigr\rvert_{t=0} &=
\Tr_{\rm S}\left(\frac{-{\rm i}}{\hbar}[H_{\rm sb}, \hat\rho(0)]\right) \nonumber\\
&= \pm g \hat\sigma_y \sum_{\alpha=0}^\infty \sqrt{\alpha+1}\,
\bigl(c_{\alpha+1}c_\alpha^* + c_{\alpha+1}^*c_\alpha\bigr) \nonumber\\
&= \pm 2g \hat\sigma_y \sum_{\alpha=0}^\infty \sqrt{\alpha+1}\,{\rm Re}(c_{\alpha+1}c_\alpha^*).
\end{align}

\noindent
It implies, in particular, for the purity that

\begin{align} \label{sbinispinrhoop2ddt}
\frac{{\rm d}}{{\rm d}t} \Tr\bigl[(\hat\rho_{\rm S}(t)\bigr]^2\Bigr\rvert_{t=0}
&= \Tr_{\rm S}\bigl[\dot{\hat\rho}_{\rm S}(0)\hat\rho_{\rm S}(0) +
\hat\rho_{\rm S}(0)\dot{\hat\rho}_{\rm S}(0)\bigr] \nonumber\\
&= \pm g \sum_{\alpha=0}^\infty \sqrt{\alpha+1}\,{\rm Re}(c_{\alpha+1}c_\alpha^*
\Tr_{\rm S}\left[\bigl(I_0 \pm \hat\sigma_x\bigr)\hat\sigma_y +
\hat\sigma_y\bigl(I_0 \pm \hat\sigma_x\bigr)\right] \nonumber\\
&=  \pm g \sum_{\alpha=0}^\infty \sqrt{\alpha+1}\,{\rm Re}(c_{\alpha+1}c_\alpha^*
\Tr_{\rm S}\left[\hat\sigma_y + {\rm i}\sigma_z + \hat\sigma_y -  {\rm i}\sigma_z\right] = 0.
\end{align}

\noindent
Defining the polarization as the vertical component of the Bloch vector,

\begin{equation} \label{sbspinpol}
a_z(t) = \langle \hat\sigma_z \rangle = \Tr_{\rm S}\bigl[\hat\sigma_z \hat\rho_{\rm S}(t)\bigr] =
\frac{1}{2} \bigl(\dot{\rho}_{\uparrow\uparrow}(t) -
\dot{\rho}_{\downarrow\downarrow}(t)\bigr)
\end{equation}

\noindent
its first time derivative at $t = 0$ is obtained as

\begin{equation} \label{sbspinpolddt}
\dot{a}_z(t) = \pm 2g \sum_{\alpha=0}^\infty \sqrt{\alpha+1}\,{\rm Re}(c_{\alpha+1}c_\alpha^*)
\Tr_{\rm S}\bigl[\hat\sigma_z \hat\sigma_y \bigr] = 0.
\end{equation}

\noindent
Along the same lines as in Eq.\ (\ref{sbinispinrhoopddt}), the initial second time
derivative of the reduced density operator is found to be

\begin{align} \label{sbinispinrhoopd2dt2}
\frac{{\rm d^2}}{{\rm d}t^2}\, \hat\rho_{\rm S}(t)\Bigr\rvert_{t=0} =&
\Tr_{\rm S}\left(\frac{-{\rm i}}{\hbar}[H_{\rm sb}, \dot{\hat\rho}(0)]\right) \nonumber\\
= &\pm 2g \sum_{\alpha=0}^\infty \sqrt{\alpha+1}\,\left(
\omega_0\hat\sigma_z {\rm Re}(c_{\alpha+1}c_\alpha^*) +
\omega_1\hat\sigma_y {\rm Im}(c_{\alpha+1}c_\alpha^*) \right) \nonumber\\
&\mp 2g^2\hat\sigma_x \sum_{\alpha=0}^\infty \left(|c_\alpha|^2(2\alpha+1) +
\sqrt{(\alpha+1)(\alpha+1)}\,{\rm Re}(c_{\alpha+2}c_\alpha^*)\right),
\end{align}

\noindent
The second time derivative of the purity reads

\begin{align} \label{sbinispinrhoop2d2dt2}
\frac{{\rm d^2}}{{\rm d}t^2} \Tr\bigl[(\hat\rho_{\rm S}(t)\bigr]^2\Bigr\rvert_{t=0} =
4g^2 \Biggl[&\left(
\sum_{\alpha=0}^\infty \sqrt{\alpha+1}\,{\rm Re}(c_{\alpha+1}c_\alpha^*)\right)^2 \nonumber\\
& -\sum_{\alpha=0}^\infty \Bigl((2\alpha+1)|c_\alpha |^2 +
2\sqrt{(\alpha+1)(\alpha+2)}\,{\rm Re}(c_{\alpha+2}c_\alpha^*)\Bigr)\Biggr],
\end{align}

\noindent
and the second time derivative of the polarization is

\begin{equation} \label{sbspinpold2dt2}
\ddot{a}_z(t) = \frac{1}{2} \bigl(\ddot{\hat\rho}_{\uparrow\uparrow}(t) -
\ddot{\hat\rho}_{\downarrow\downarrow}(t)\bigr) =
\pm 2g \omega_0 \sum_{\alpha=0}^\infty \sqrt{\alpha+1}\,{\rm Re}(c_{\alpha+1}c_\alpha^*).
\end{equation}

\reftitle{References}
\externalbibliography{no}

\end{document}